\documentclass[fleqn,usenatbib]{mnras}

\usepackage{newtxtext,newtxmath}

\usepackage[T1]{fontenc}

\DeclareRobustCommand{\VAN}[3]{#2}
\let\VANthebibliography\thebibliography
\def\thebibliography{\DeclareRobustCommand{\VAN}[3]{##3}\VANthebibliography}

\usepackage{graphicx}	
\usepackage{amsmath,verbatim}	

\title[Dust dynamics in GI discs]{Dust trapping, collisional velocity and velocity dispersion in gravito-turbulent discs}

\author[C. Longarini et al.]{
Cristiano Longarini$^{1}$\thanks{E-mail: cl2000@cam.ac.uk}, Cathie J. Clarke$^{1}$, Richard A. Booth$^{2}$ and Cat Leedham$^{1}$ 
\\
$^{1}$Institute of Astronomy, University of Cambridge, Madingley Road, Cambridge, CB3 0HA, United Kingdom\\
$^{2}$School of Physics and Astronomy, University of Leeds, Leeds LS2 9JT, United Kingdom
}

\date{Accepted 17th of August 2026. Received 12th of August; in original form 3rd of June}
    
\pubyear{\the\year{}}

\begin{document}
\label{firstpage}
\pagerange{\pageref{firstpage}--\pageref{lastpage}}
\maketitle

\begin{abstract}
High-resolution ALMA observations indicate that planet formation is already well underway in the earliest stages of disc evolution, when discs are typically massive enough to be regulated by gravitational instability (GI). In this regime, the interplay between dust dynamics and GI may enable rapid core formation, but the conditions under which this process operates remain poorly constrained. In this work, we use three-dimensional global SPH simulations to investigate the dynamics of dust particles in gravitationally unstable discs over a wide range of Stokes numbers. We focus on three key quantities that regulate early planet formation: dust trapping in spiral arms, collisional velocities between dust grains, and the dust velocity dispersion. We find that particles with Stokes numbers in the range $\mathrm{St}\simeq0.1$–$1$ undergo the strongest concentration within spiral arms while simultaneously exhibiting low velocity dispersion. This combination makes this aerodynamic regime the most favourable for the onset of dust-driven gravitational instability and direct dust collapse. Where direct comparisons of collision velocities are possible, our three-dimensional results are in excellent agreement with previous two-dimensional studies and are consistent with a picture in which grains are partially coupled to a Kolmogorov-like turbulent velocity field in the gas. Our results indicate that, in young self-gravitating discs, dust particles in this intermediate coupling regime provide a natural pathway to the formation of planetary cores. 
\end{abstract}




\section{Introduction}
The ubiquity of substructures in Class II protoplanetary discs \citep{andrews18} has opened a debate on how planet formation timescales relate to the evolutionary stage of young stellar objects. If these substructures are interpreted as signatures of embedded planets, a robust conclusion is that a substantial fraction of planet formation must occur while protostellar discs are still young and massive. Consistent with this picture, observational signposts of planets have been reported in very young discs, with ages below 500 kyr \citep{sheehan18, segura-cox20}.

These findings indicate that planet formation must occur within the first Myr of disc evolution, posing a challenge to standard planet formation scenarios such as core accretion \citep{pollack96}. At such early stages, protoplanetary discs are expected to be massive, as a large fraction of their mass has not yet been accreted onto the central object, making disc self-gravity dynamically important. A natural consequence is the onset of gravitational instabilities, which give rise to large-scale spiral structures that transport angular momentum and strongly influence the dynamics of both gas and dust.

Historically, gravitational instability (GI) has been proposed as a formation pathway for giant planets \citep{boss97}. However, this scenario has fallen out of favour, as it is generally expected to preferentially produce stellar or sub-stellar companions rather than planets \citep{kratter16}. Indeed, the typical initial mass of fragments formed through GI is of order $1$–$10,\mathrm{M_{Jup}}$. Such objects are then expected to undergo rapid gas accretion from the surrounding disc, likely growing into stellar companions.

The interplay between dust dynamics and gravitational instability has been shown to provide a viable pathway for the formation of planetary cores in young protostellar discs \citep{rice04,rice06}. In this framework, dust grains are efficiently trapped within gaseous spiral arms, where their concentration can become sufficiently high to trigger the direct gravitational collapse of the dust component, leading to the formation of planetary cores. This scenario has been extensively investigated over the past decade \citep{booth16,gibbons12,gibbons14,gibbons15,baehr21}, demonstrating that, for sufficiently large dust particles, direct collapse can indeed occur \citep{baehr22,longarini23b,rowther24,rice25}. A key parameter in this process is the Stokes number, which quantifies the degree of aerodynamic coupling between gas and dust. 

In this work, we investigate the dynamics of solid particles in gravitationally unstable discs across different aerodynamic regimes using three-dimensional global SPH simulations. The paper is structured as follows. In Section~\ref{constst}, we describe the numerical framework and simulation setup. In Section~\ref{sec_analysis}, we present the analysis of the simulations, focusing on three key quantities: the dust-to-gas ratio, the collisional velocity, and the velocity dispersion. In Section~\ref{sec_discussion}, we discuss our results in the context of dust growth and dust collapse in GI discs. Finally, in Section~\ref{sec_concl}, we summarise our main conclusions.

\begin{figure*}
    \centering
    \includegraphics[width=1\linewidth]{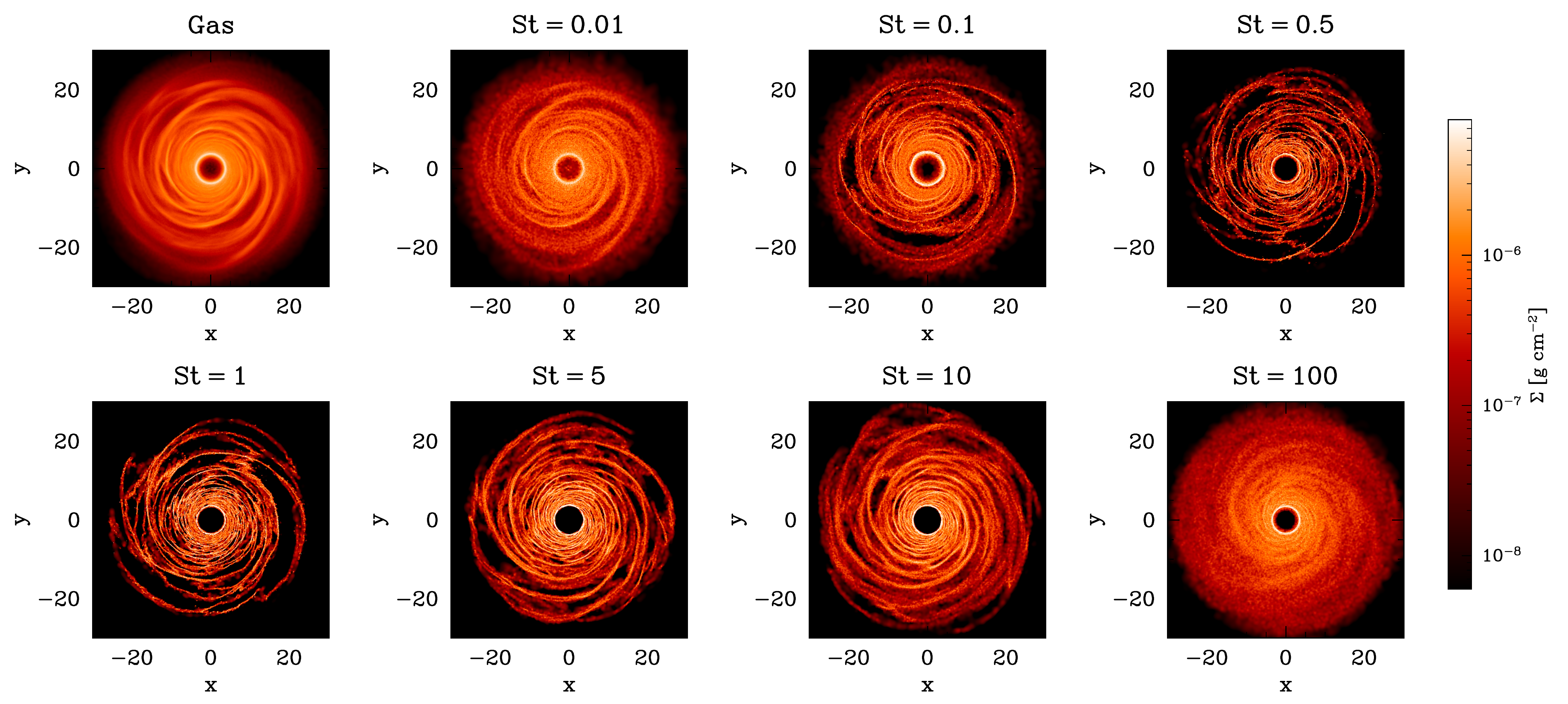}
    \caption{ Surface density of gas (top left panel) and dust particles, for different Stokes numbers: in order $\{0.01,0.1,0.5,1,5,10,100\}$.}
    \label{fig:collection_sims}
\end{figure*}

\section{Methods}\label{constst}

In this work, we perform numerical SPH simulations of gas and dust protostellar discs using the code \textsc{phantom} \citep{phantom}. This code is widely used in the astrophysical community to study gas and dust dynamics in accretion discs \citep{ceppi23, prasad25}, both in a single fluid mixture \citep{nealon20} or dust-as-particles approach \citep{aly24}. In this work, we use the dust-as-particles formulation.

 {\subsection{Two-fluid gas and dust mixtures}}

 {The two-fluid formulation in \textsc{Phantom} is based on the continuum equations in the form
\begin{equation}
    \frac{\partial \rho_{\mathrm{g}}}{\partial t} + (\boldsymbol{v}_{\mathrm{g}} \cdot \nabla)\rho_{\mathrm{g}} = -\rho_{\mathrm{g}}(\nabla \cdot \boldsymbol{v}_{\mathrm{g}}),
\end{equation}
\begin{equation}
    \frac{\partial \rho_{\mathrm{d}}}{\partial t} + (\boldsymbol{v}_{\mathrm{d}} \cdot \nabla)\rho_{\mathrm{d}} = -\rho_{\mathrm{d}}(\nabla \cdot \boldsymbol{v}_{\mathrm{d}}),
\end{equation}
\begin{equation}
    \frac{\partial \boldsymbol{v}_{\mathrm{g}}}{\partial t} + (\boldsymbol{v}_{\mathrm{g}} \cdot \nabla)\boldsymbol{v}_{\mathrm{g}} = -\frac{\nabla P}{\rho_{\mathrm{g}}} + \frac{\rho_d}{\rho_{\mathrm{g}}}\frac{1}{(\rho_g+\rho_d)t_s}(\boldsymbol{v}_{\mathrm{d}} - \boldsymbol{v}_{\mathrm{g}})- \nabla (\Phi_\star + \Phi_g),
\end{equation}
\begin{equation}
    \frac{\partial \boldsymbol{v}_{\mathrm{d}}}{\partial t} + (\boldsymbol{v}_{\mathrm{d}} \cdot \nabla)\boldsymbol{v}_{\mathrm{d}} = -\frac{1}{(\rho_g+\rho_d)t_s}(\boldsymbol{v}_{\mathrm{d}} - \boldsymbol{v}_{\mathrm{g}})- \nabla (\Phi_\star + \Phi_d),
\end{equation}
where $\mathbf{v}_{\mathrm{g,d}}$ are the gas and dust velocity field, $t_s$ is the stopping time, $\Phi_\star$ is the stellar gravitational potential, $\Phi_g$ is the gas self-gravity and $\Phi_d$ the dust one. The drag term couples the gas to the local dust flow. The previous equations are discretised following eqs. 239-240 of \citet{phantom}, with the gas and dust densities computed by kernel summation over particles of the same type, 
}
\begin{equation}\label{eq:rhogas}
    \rho_{a}=\sum_{b} m_{b} W_{ab}\left(h_{a}\right), \quad 
    h_{a}=h_{\mathrm{fact}}\left(\frac{m_{a}}{\rho_{a}}\right)^{1/3},
\end{equation}
\begin{equation}\label{eq:rhodust}
    \rho_{i}=\sum_{j} m_{j} W_{ij}\left(h_{i}\right), \quad 
    h_{i}=h_{\mathrm{fact}}\left(\frac{m_{i}}{\rho_{i}}\right)^{1/3},
\end{equation}
where subscripts $a,b,c$ refer to gas particles and $i,j,k$ to dust particles \citep{monaghan2fl}, $W$ is the SPH smoothing kernel, $h$ is the smoothing length, and $h_\text{fact}=1/2$. Following \citet{monaghan2fl}, the gas-dust drag interaction is evaluated using a double-hump kernel, which vanishes at zero separation and peaks around one smoothing length; together with the symmetric drag formulation of equations~(2)--(3), this guarantees exact conservation of linear and angular momentum.

It is worth clarifying in what sense the dust component is treated as a fluid in this scheme. Dust particles are not `superparticles' in the particle-in-cell sense of grid-based codes: they are interpolation points sampling an underlying continuous field, and the algorithm treats gas and dust symmetrically, differing only in that the dust phase is pressureless. 
At the same time, the dust equations of motion do not fundamentally presuppose this fluid closure. In the limit $t_s\to\infty$, the drag coupling to neighbouring gas particles becomes negligible and the dust particles evolve as independent, freely streaming bodies, indistinguishable from an N-body system. More generally, for standard Epstein or Stokes drag laws (eqs.~240, 250 and 254 of \citealt{phantom}), the total gas+dust density cancels from the expression for $t_s$, so that the acceleration of an individual dust particle depends only on the interpolated gas velocity at its position (and on gravity), not directly on the local particle density. The double-hump-kernel density summation in equation~(\ref{eq:rhodust}) therefore enters the scheme mainly through the gas back-reaction and through the definition of $t_s$, rather than being a necessary ingredient of the dust particle's own equation of motion. The fluid picture is thus a good description of the scheme's behaviour in the well-coupled, low-Stokes-number regime, but the underlying algorithm solves a more general set of equations that smoothly recovers independent-particle (N-body) dynamics at high Stokes number -- which is relevant to the trend in collision velocities discussed below.

In this work, we evolve dust particles with a fixed Stokes number throughout the disc, defined as
\begin{equation}
    \text{St} = t_s \Omega_k,
\end{equation}
where $\Omega_k$ is the Keplerian orbital frequency. The drag algorithm in \textsc{Phantom} computes the stopping time, and evolves the particles accordingly. Hence, to fix the Stokes number throughout the simulation, we compute the stopping time for the $i$-th particle as  {
\begin{equation}
    t_{s,i}(R_i) = \text{St}\sqrt{\frac{R_i^3}{G M_\star}}.
\end{equation}}

We found that running simulations at fixed Stokes number significantly improves numerical feasibility. In the dust-as-particles approach, fixing the grain size leads to large spatial variations in the Stokes number causing the code to struggle in resolving the physics across both short and long timescales. In particular, this algorithm becomes inefficient for small particles, i.e. for dust populations whose Stokes number approaches  {zero}. By contrast, enforcing a constant Stokes number maintains a fixed ratio between stopping and dynamical times throughout the disc, allowing us to reliably explore small Stokes numbers (down to St $\sim10^{-2}$ in this case). In these simulations, we are using the velocity reconstruction procedure presented in \cite{price20}.

\subsection{Heating and cooling}
To account for the effect of cooling or heating phenomena, we write the complete equation for the evolution of gas internal energy $e$
\begin{equation}
    \frac{\partial e}{\partial t}+\left(\mathbf{v}_{\mathrm{g}} \cdot \nabla\right) e =-\frac{P}{\rho_{\mathrm{g}}}\left(\nabla \cdot \mathbf{v}_{\mathrm{g}}\right)+\Lambda_\text{shock} - \frac{\Lambda_\text{cool}}{\rho_g}+\frac{\Lambda_{\mathrm{drag}}}{\rho_g},
\end{equation}
where the first term on the RHS is the $P\text{d}V$ work, the second is a heating term due to the shock viscosity, the third is the cooling of the disc and the last term is the drag heating term. In this work, we assume an adiabatic equation of state. For an ideal gas, it is possible to link pressure and density as follows
\begin{equation}
    P = (\gamma-1)\rho_g e = \frac{c_s^2\rho_g}{\gamma},
\end{equation}
where $\gamma = 5/3$ and $c_s$ is the adiabatic sound speed, that is initialized as a power law $c_s\propto R^{-0.25}$. 

The shock viscosity term in \textsc{phantom} \citep{phantom} takes the form
\begin{equation}
    \Lambda_\text{shock} \propto \alpha^\text{AV} \rho c_s h |\nabla \cdot \mathbf{v}| 
    + \beta^\text{AV} \rho h^2 |\nabla \cdot \mathbf{v}|^2,
\end{equation}
where $\alpha^\text{AV}$ and $\beta^\text{AV}$ are the linear and quadratic 
viscosity coefficients. In a stratified disc, the velocity divergence 
associated with turbulent fluctuations scales as 
$|\nabla \cdot \mathbf{v}| \sim \delta v / h \sim c_s / H$, so that 
the two terms scale respectively as $\alpha^\text{AV}(h/H)$ and 
$\beta^\text{AV}(h/H)^2$, recovering the resolution-dependent 
form discussed in the literature. The viscosity term is dissipative 
and heats the disc. In the \textsc{phantom} simulations we are not using the \textsc{disc-viscosity} flag, meaning that the shock capturing viscosity is not described by an $\alpha_\text{SS}$  prescription. We did so since in these systems the main driver of angular momentum transport is GI.

For the cooling we use the prescription from \cite{gammie01} and \cite{rice04}, in which the cooling time $t_\text{cool}$ is proportional to the dynamical time, with a factor of proportionality $\beta_\text{cool}$
\begin{equation}
    t_\text{cool} = \beta_\text{cool}\Omega^{-1}
\end{equation}
Under the assumption that
the transfer of angular momentum driven by gravito-turbulence occurs locally \citep{Lodato04,Bethune21}, we can relate the cooling parameter to an effective $\alpha-$viscosity parameter
\begin{equation}\label{alphacool}
    \alpha_\text{GI} = \frac{4}{9}\frac{1}{\gamma(\gamma-1)\beta_\text{cool}}.
\end{equation}
The drag heating term for an individual dust particle $i$ is
\begin{equation}
    \Lambda_{\mathrm{drag},i} = \frac{\rho_g\rho_d}{(\rho_g+\rho_d)t_s}
    |\mathbf{v}_{d,i}-\mathbf{v}_{\mathrm{g}}|^2,
\end{equation}
where $\mathbf{v}_{d,i}$ is the velocity of particle $i$. This quantity  is computed per-particle and then summed to give the total drag heating,  naturally accounting for the full distribution of dust velocities  rather than just the mean flow.

\subsection{Simulations' setup}
We initially evolve a 1 million particles gas-only disc with a mass $M_d=0.1\text{M}_\odot$ around a solar mass star, modelled as a sink particle. The inner radius of the disc is $R_{\rm in}=0.25$ and the outer one $R_{\rm out}=25$ in code units. The initial aspect ratio is chosen so that the minimum Toomre parameter $Q$ at the outer radius is 2. To trigger gravitational instability we cool the disc using the $\beta-$cooling prescription, where $\beta$ is the ratio between the cooling time and the dynamical time, and we choose $\beta=10$, to ensure that the disc does not fragment \citep{deng17}. The shock capturing viscosity coefficients used in the simulations are $\alpha_{\rm AV}\in[0,0.1]$ and $\beta_{\rm AV} =2$ {, and we used the Cullen and Dehnen switches \citep{cullen10}}. After an outer thermal time, i.e. 10 dynamical outer orbits, we add dust with a global initial dust to gas ratio of $10^{-2}$. For the dust-as-particles scheme, we adopt a gas-to-dust particle ratio of 5. As described above, the dust is evolved to maintain a constant global Stokes number. Simulations including dust are evolved for 10 outer stopping times for St$\leq1$, while for higher Stokes numbers they are evolved for 100 orbital periods. 

With $N_g = 10^6$ gas particles, the average unperturbed $\langle h/H \rangle_g = 0.25$, where this quantity indicates how many particles resolve the vertical scale height. This value is safe, ensuring good resolution for the shock-capturing viscosity. As for the dust, $\langle h/H \rangle_d = 5^{1/3}\langle h/H \rangle_g$, since the ratio of the number of particles is 5. This value still ensures that the Jeans length in the dust is resolved, as discussed in \citet{longarini23b}. We also perform a gas-only simulation with $N=1.2\times10^7$ particles, which results in a $\langle h/H\rangle = 0.1$.  {Table \ref{table1} describes the simulations run in this work.}

\begin{table}\caption{List of the simulations performed in this work}
\begin{tabular}{llll}
Simulation & Stokes number  & Running time $[\Omega_{\rm out}^{-1}]$ & \\
\hline 
\textbf{S1}   &  0.01&   1 & \\
\textbf{S2}     &  0.1&  10  & \\
\textbf{S3}  &  0.5&   50 & \\
\textbf{S4}   &  1&   100 & \\
 \textbf{S5}   &  5&   100 & \\
 \textbf{S6}    &  10&  100 & \\
 \textbf{S7}     &  100&  100 & \\
\end{tabular}
\label{table1}
\end{table}

\section{Analysis and results}\label{sec_analysis}

\subsection{Power spectrum and smallest resolvable scale}\label{sec:pspec}
\begin{figure}
    \centering
    \includegraphics[width=\linewidth]{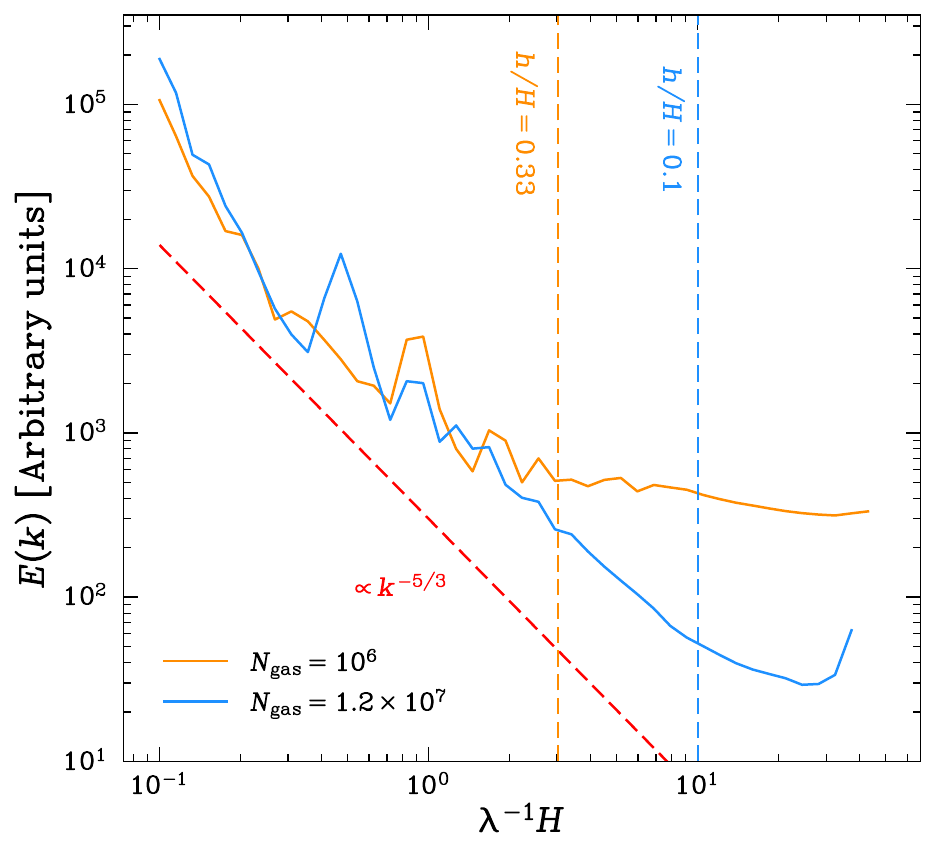}   
    \caption{Power spectrum of the turbulent kinetic energy, $E(k)$, as a function of $\lambda^{-1}H$ for two different gas resolutions,  {where $\lambda = 2\pi /k$}. In both cases, the spectrum follows a Kolmogorov scaling, $E(k) \propto k^{-5/3}$, over a broad range of spatial scales before flattening at small scales due to numerical noise associated with inter-particle jitter. The lower-resolution run $(N=10^6)$, which corresponds to the resolution adopted in the simulations analysed in this work, already provides a satisfactory resolution of the turbulent cascade down to scales comparable to the mean smoothing length.}
    \label{pspectrum}
\end{figure}

Before discussing the dynamics of dust particles in gravito-turbulent discs, we first assess whether gas turbulence is adequately resolved in our simulations. \citet{booth19} characterised gravito-turbulence using local shearing-box simulations, showing that the turbulent energy cascade follows a Kolmogorov scaling with wavenumber, $E(k) \propto k^{-5/3}$, down to a scale determined by the numerical resolution. We hence compute the power spectrum of the turbulent motions in our gas simulation, before adding dust particles, according to the following procedure.

We compute the velocity field of the turbulent fluctuations  after subtracting the axisymmetric mean flow from the particle velocities. The disc is decomposed into a set of narrow radial annuli, allowing the analysis to be performed in regions that can be approximated as locally homogeneous.

Within each annulus, the local disc scale height is estimated from the sound speed, assuming hydrostatic equilibrium. The disc is decomposed into $N_{\rm ann} = 100$ radial annuli of  half-width $\delta R = 0.5$ (in code units). The residual velocity field is then interpolated onto a uniform  three-dimensional Cartesian grid, and a three-dimensional  Fourier transform is performed to obtain the velocity field in Fourier space. The energy density is defined as
\begin{equation}
    E({k}) =
\frac{1}{2}
\left(
|\delta v_x|^2 +
|\delta v_y|^2 +
|\delta v_z|^2
\right),
\end{equation}
from which the isotropic power spectrum $E(k)$ is obtained by averaging  over spherical shells in wavenumber space. For each annulus, spatial  scales are expressed in dimensionless form using the local scale height,  defining the wavenumber $kH(R)$. This procedure yields a set of spectra  $E(k)$ computed at different disc radii, each containing at least  $N_{\rm min} = 1000$ gas particles. We find that the spectra obtained  at different radii are mutually consistent within the overlapping range  of scales. They are therefore combined onto a common  {$kH/(2\pi)$ grid of 50  logarithmically spaced bins in the range $kH/(2\pi) \in [0.1, 100]$}, and  averaged using a geometric mean to construct a single radially averaged  power spectrum, which accounts for the radial variation of the disc  scale height.

Figure~\ref{pspectrum} shows the turbulent energy spectrum at two different resolutions, to show its impact on the dissipation scale,  {where $\lambda = 2\pi /k$}. The standard resolution ($N_g=10^6$) is the orange line, while the high resolution test ($N_g=1.2\times10^7$) is the blue line. The figure shows that the turbulence spectrum follows a Kolmogorov scaling, $\propto k^{-5/3}$, down to a characteristic scale below which it flattens as a result of inter-particle jitter. This scale therefore represents the minimum scale at which turbulence in the gas is properly resolved, and is consistent with the average value of $h$ for the gas particles. 

This picture is consistent with the findings of \cite{booth19}, who investigated the properties of turbulence in 3D shearing-box simulations. Despite the global nature of our simulations, our standard-resolution run ($N_g = 10^6$) resolves turbulence down to  {approximately $\lambda =  0.1-0.3H$}, comparable to the lower-resolution runs of \cite{booth19}, corresponding to 8 cells per scale height. This agreement is reasonable given the global nature of our simulations.

 {We note that the degree to which SPH can faithfully represent subsonic Kolmogorov turbulence depends not only on the dissipation scale itself, but also on the numerical Reynolds number. In finite-volume grid codes, numerical dissipation is implicit and linear in the velocity, so the numerical Re is independent of the Mach number. In SPH, by contrast, the numerical viscosity is explicit and resolution-dependent, so the numerical Re has an explicit dependence on the Mach number, becoming progressively harder to resolve at lower Mach number \citep{price12}. As a result, increasing the SPH resolution acts on turbulence in two ways: it pushes the dissipation scale to smaller $kH$, and it simultaneously reduces the numerical viscosity, allowing a higher Re to be resolved and the inertial range to be more faithfully reproduced. This may explain why the standard-resolution spectrum appears to depart from the Kolmogorov scaling earlier than expected, while the high-resolution run more closely reproduces the expected inertial-range behaviour down to smaller scales.}

Remarkably, in the high-resolution SPH run ($N_g = 1.2 \times 10^7$, $h/H\sim0.1$), the minimum resolved scale is comparable to that achieved in high-resolution shearing-box simulations (32–64 cells per scale height). This suggests that SPH approaches can be highly effective at resolving turbulence. However, the global nature of our setup introduces additional computational costs. We therefore speculate that an SPH shearing-box configuration could provide an optimal compromise, combining the efficiency of SPH in resolving turbulent fluctuations with the reduced computational expense of a local framework.

\subsection{Implications for dust collision velocity}
The demonstration that our simulations generate a power spectrum consistent with Kolmogorov turbulence has direct implications for the expected dependence of dust collision velocities on the Stokes number. In a turbulent flow with a Kolmogorov cascade, the velocity difference across a scale $\ell$ scales as $\delta v(\ell) \propto \ell^{1/3}$, while the eddy turnover time scales as $t_{\ell} \sim \ell / \delta v(\ell) \propto \ell^{2/3}$. Eliminating $\ell$, the turbulent velocity at a given turnover time is
\begin{equation}
    \delta v \propto t_\ell^{1/2}.
\end{equation}
The condition $t_\ell \sim t_s$ is critical because eddies with $t_\ell \gg t_s$ advect particles coherently (contributing little to their relative velocities), while eddies with $t_\ell \ll t_s$ are too  {weak} to significantly perturb particle motions before drag damps them.

Treating the velocity kicks from eddies as a random walk: each eddy 
of turnover time $t_\ell$ imparts an acceleration of amplitude 
$\delta v(\ell)/t_\ell$ on the particle for a correlation time $t_\ell$, 
but the particle response is suppressed by drag over the stopping time $t_s$. 
The contribution to the mean-square velocity from eddies with $t_\ell < t_s$ is {
\begin{equation}
    \langle \Delta v^2 \rangle_\ell \sim 
    \left(\frac{\delta v(\ell)}{t_\ell}\right)^2 t_\ell \cdot t_s 
    \propto \frac{t_\ell^2}{t_s},
\end{equation}}
which increases with $t_\ell$ and is therefore dominated by the largest 
eddies satisfying $t_\ell \lesssim t_s$. Eddies with $t_\ell > t_s$ 
introduce correlated motions and can be neglected. Substituting 
$t_\ell \sim t_s$ directly gives $\langle \Delta v^2 \rangle \propto t_s$, 
independent of $\ell$, hence \citep{volk80, ormel07, dubrulle95}
\begin{equation}
    \Delta v \sim \delta v(t_s) \propto t_s^{1/2}.
\end{equation}
Since $\text{St} = t_s \Omega$, this leads to the expected scaling $\Delta v \propto \text{St}^{1/2}$. We underline that this is valid when the largest eddy turnover time is $>t_s$, and the turnover time of the fastest one is $<t_s$. In Section~\ref{res_coll}, we will consider the extent to which the measured collision velocities are consistent with this $\text{St}^{1/2}$ scaling and identify the numerical features which cause a deviation from this relation at small Stokes numbers.

\subsection{The competition of two effects}
When studying the dynamics of dust particles in gravitationally unstable discs, there are two main phenomena to take into account: \textbf{dust trapping} in gas spiral arms and the development of relative motions between dust particles, which we call \textbf{dust excitation}. 

Gas spiral arms act as gravitational potential minima, and it is this property that drives dust concentration within them. While pressure forces resist the compression of gas, dust particles are collisionless and do not experience them: they therefore sink into the potential well unimpeded, driven by gravity and mediated by drag. This asymmetry between gas and dust response is the same physics underlying radial drift in a smooth disc, but here operating azimuthally towards the spiral rather than radially towards the star. The combined effect can significantly increase the dust-to-gas ratio, reaching values of order unity \citep{rice04,dipierro15}.

Dust trapping is particularly efficient in gravitationally unstable spirals because the difference of speed between the spiral density wave and the background disc is small. Indeed, when gravitational instability is driven by cooling, the pattern speed $\Omega_p \sim \Omega$, as shown in \citet{cossins09}, and at zeroth order we expect the dust to follow the same behaviour. Conversely,  {in the case where the spiral arms are triggered by the presence of a planet}, where the pattern speed is constant, dust trapping is not particularly efficient, since the difference of speed between the spiral wave and the background disc increases far from the planet. 

The interaction between gas spiral arms and dust excites relative motions between solid particles. In this context, there are two important quantities to take into account: the collisional velocity and the velocity dispersion. The collisional velocity is the relative velocity between two dust grains at zero separation and is a key quantity for dust growth. Laboratory experiments and dust evolution models show that collisions at velocities exceeding the fragmentation threshold lead to destructive outcomes rather than growth (e.g. \citealt{birnstiel16, birnstiel24}). For compact silicate aggregates, the fragmentation velocity is of order $v_{\rm frag} \sim 1\,\mathrm{m\,s^{-1}}$ \citep{guttler10}, while for icy aggregates this threshold has traditionally been considered significantly higher, typically $v_{\rm frag} \sim 10\,\mathrm{m\,s^{-1}}$ \citep{gundlach15}. However, this picture has been complicated by more recent laboratory experiments. \citet{gartner17} showed that the surface energy of water ice increases significantly above $\sim 200\,\mathrm{K}$, suggesting that the enhanced stickiness of ice is only effective in a relatively narrow temperature range. Furthermore, \citet{musiolik19} measured the sticking and rolling properties of mm-sized ice grains at different temperatures and found that below $\sim 175\,\mathrm{K}$ the surface energy of water ice is comparable to that of silicate dust, implying that ice grains do not have a significant sticking advantage over silicates at the low temperatures typical of the outer disc. Taken together, these results suggest that the fragmentation threshold for icy aggregates may be closer to the silicate value in much of the disc, and that the exact value of $v_{\rm frag}$ remains uncertain. Collisions above these velocities therefore hinder grain growth and limit the maximum grain size attainable. 

Conversely, the velocity dispersion is the statistical measure of the spread of velocities in a system, defined as the standard deviation of the velocity distribution on a given scale. This quantity can be interpreted as an effective sound speed, even though the origin of such dispersion is not related to thermal processes. Its value determines the stability of the dust layer against gravitational collapse, as discussed in \citet{longarini23a}.

These two effects, dust trapping and dust excitation, are in competition, and their interplay determines the dynamics of dust particles. Dust trapping tends to increase dust density in spiral arms, making these locations suitable for dust growth and, possibly, gravitational collapse. At the same time, the development of relative motions tends to stabilise the dust layer, and hinders the process of growth.

\subsection{Dust trapping}
Dust trapping is very efficient in gas spiral arms \citep{dipierro15,rice04,longarini23b,rowther24}. This occurs because spiral arms correspond to pressure maxima, and as a consequence of the aerodynamic coupling, dust particles tend to accumulate in them. The strength of dust trapping depends on the Stokes number, which sets the degree of coupling between gas and dust. We expect that for small Stokes numbers (St < 0.1) dust particles essentially follow the gas motion, and trapping is not efficient. In this regime, particles are too strongly coupled and simply trace the gas density without any significant enhancement. Additionally, the dust settling time is too long compared to  {the spiral arms} lifetime, making this process negligible. For Stokes numbers close to unity (St $\in [0.1, 1]$) we expect the strongest dust trapping: here the stopping time is comparable to the dynamical time, leading to the largest dust-density enhancements. Finally, for larger Stokes numbers, the drag force becomes too weak to dissipate the relative velocity between dust and gas quickly enough for the dust to become trapped in the potential minimum of the spiral arms, and dust trapping no longer occurs.

\begin{figure}
    \centering
        \includegraphics[width=\linewidth]{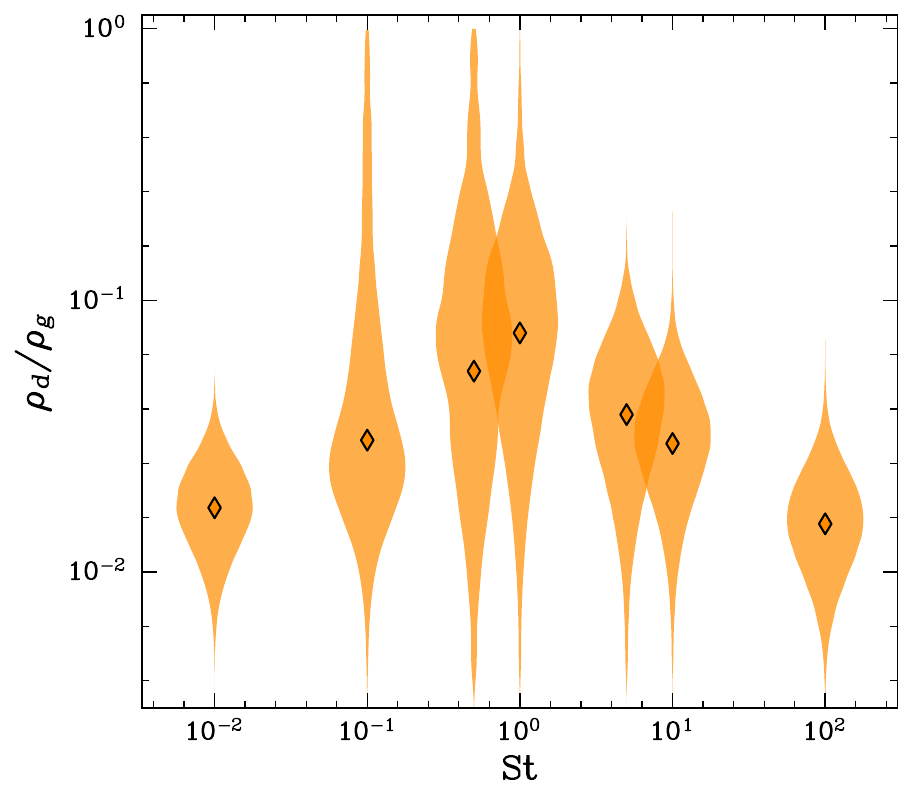}
    \caption{Violin plots of the distributions of dust to gas ratio for different Stokes number. The diamonds correspond to the median values of the distributions, showing a maximum for St$=1$. }
    \label{fig:dtgratio}
\end{figure}

This trend is evident from the different panels of Figure \ref{fig:collection_sims}, and this is shown more quantitatively in Figure \ref{fig:dtgratio}. The latter figure shows that for St $\sim0.01$ the dust-to-gas ratio distribution is packed around  $\sim 10^{-2}$. Conversely, for St $\in[0.1, 1]$, the distributions have an increasing median value, and they are much broader, covering a range up to $\sim 1$. Then, for St $>1$, the median value decreases again, because trapping is not efficient.



\subsection{Dust collision velocity}
We compute the collision velocity for a pair of dust particle following \cite{booth16}.

We choose a dust particle and we consider the particles within a smoothing sphere, corresponding to approximately $N_{\rm neigh} = 50$. In contrast to \cite{booth16}, our simulations are 3D and the local dust resolution is lower, due to numerical limitations. For this reason, the number of neighbouring particles we choose to compute the collision velocity is lower (50 against 200). For a given particle, its velocity can be described by two components, a Keplerian one and a random one
\begin{equation}
    \mathbf{v} = \mathbf{v}_{k}(R) + \delta\mathbf{v},
\end{equation}
where the Keplerian component depends on the distance of the particle from the central object. In an infinite resolution scenario, the collision velocity between two particles $a-b$ would simply be $\delta\mathbf{v}_a-\delta\mathbf{v}_b$, because their separation tends to zero. However, because of the finite resolution, two particles have a finite distance $\delta R$, and therefore their difference of speed is  {
\begin{equation}
    \mathbf{v}_i - \mathbf{v}_j = \Delta \mathbf{v}_{\rm coll, ij} + \Delta\mathbf{v}_{\rm KS}, 
\end{equation}
where $\Delta\mathbf{v}_{\rm KS}$ is the difference of velocity due to the Keplerian shear. We want to subtract this quantity to get the pure collisional velocity. To do so, we need to evaluate the gradient of the velocity field in the position of the $a-$th particle. In SPH the gradient of the velocity field can be written as \citep{phantom}
\begin{equation}
     \frac{\partial v_{i}^{\alpha}}{\partial x_{i}^\beta} = \frac{1}{\Omega_i\rho_i}\sum_j m_j (v_i-v_j)^\alpha\nabla_i ^\beta W_{ij}(h_i),
\end{equation}
where $\alpha,\beta$ runs over the velocity component and the $\Omega_a$ term is a grad-h correction that arises from the use of an adaptive smoothing length 
\begin{equation}
    \Omega_i = 1 - \frac{\partial h_i}{\partial \rho_i} \sum_j m_j \frac{\partial W_{ij}(h_i)}{\partial h_i}
\end{equation}
and
\begin{equation}
    \frac{\partial h_i}{\partial \rho_i} = -3\frac{h_i}{\rho_i}.
\end{equation}
We note that this gradient estimator differs from the one used internally by \textsc{phantom} for dust particles  {since we are using the classic bell-shaped, instead of the double-hump}; however, this difference is not expected to significantly affect the results. Hence, the Keplerian shear contribution is given by
\begin{equation}
    \Delta{v}_{\rm KS}^\alpha = \frac{\partial v_{a}^{\alpha}}{\partial x_{i}^\beta}\cdot R_{ij}^\beta.
\end{equation}
Finally, the collisional velocity between the particle $a$ and $b$ is
\begin{equation}
    \Delta v_{\rm coll, ab} = \left| \mathbf{v}_a - \mathbf{v}_b - \Delta\mathbf{v}_{\rm KS} \right|.
\end{equation}
}
To properly compute the collision velocity between dust particles, there is a resolution requirement, as discussed in \cite{booth16}. In practice, the collision velocity between the $a-$th and $b-$th particle is numerically resolved if the separation of the two particles $\Delta R_{ab}$ is smaller than the stopping distance between the two particles, defined as 
\begin{equation}
    \lambda_{\rm stop} = \Delta v t_s .
\end{equation}
Hence, the resolution condition is 
\begin{equation}
    \Delta R_{ab} < t_s \Delta v_{ab}^{\rm coll}.
\end{equation}

\begin{figure}
    \centering
    \includegraphics[width=\linewidth]{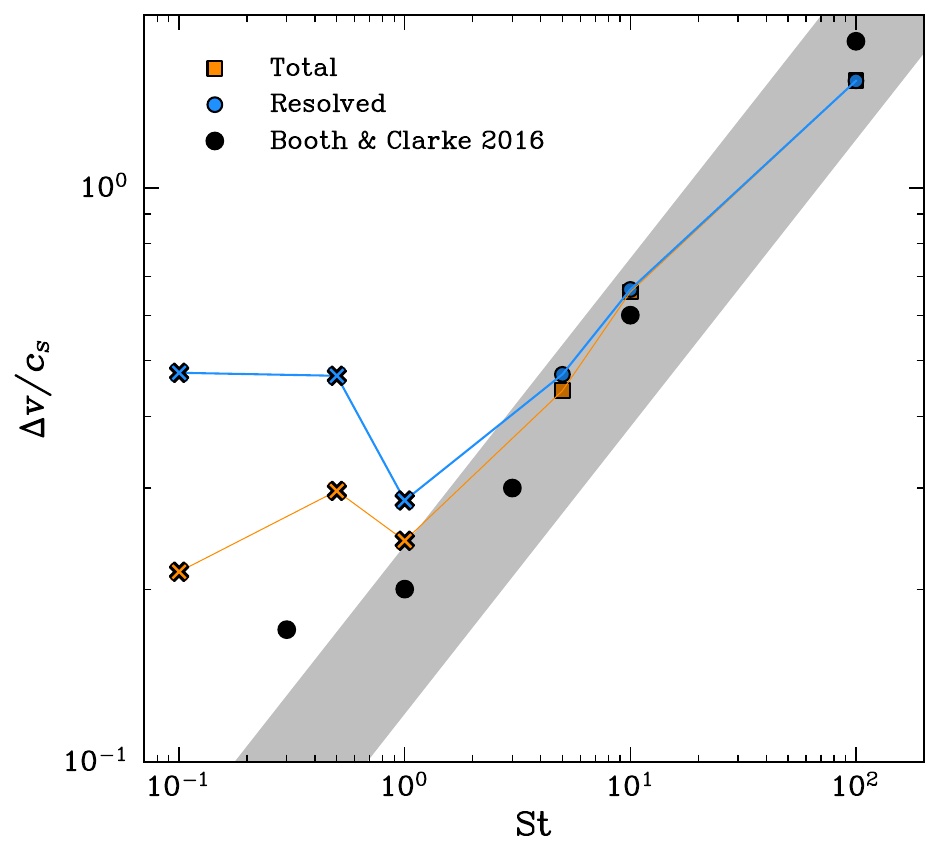}
    \caption{Median collision velocity of dust particles as a function of the Stokes number. The light blue line corresponds to the median value of resolved pairs, the orange line the median value of all pairs, and the black dots the results of \citet{booth16} for the resolved pairs at the highest resolution employed. The crosses correspond to the Stokes number regime that we are not able to properly sample because of resolution requirements for the gas (see \ref{res_coll}). The grey shaded line corresponds to $\propto \rm St^{1/2}$, that is the expected scaling of the quantity according to Kolmogorov turbulence (see \ref{res_coll}). }
    \label{collision_percentage}
\end{figure}

\begin{figure*}
    \centering
    \includegraphics[width=0.75\linewidth]{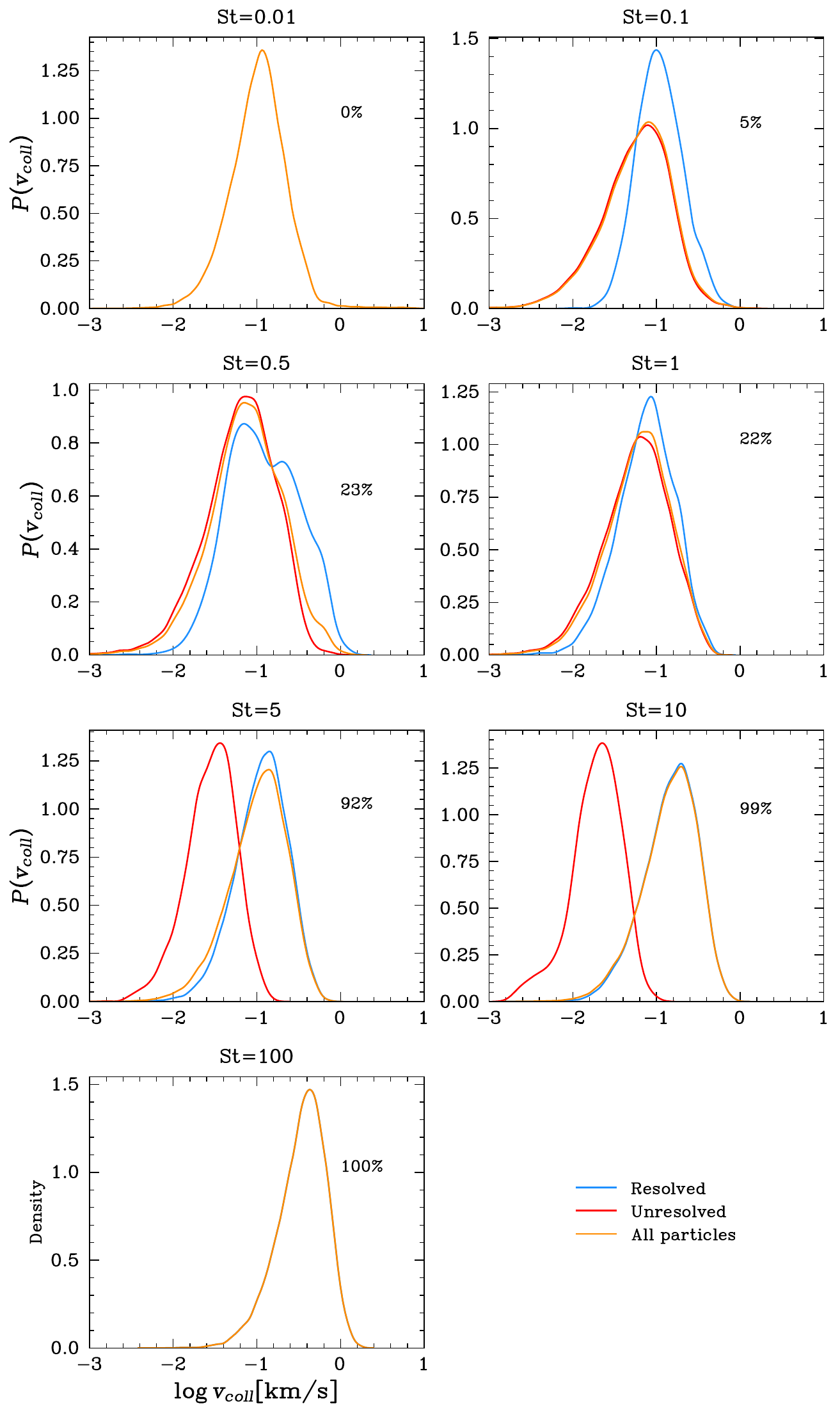}
    \caption{Distribution of the collisional velocity for the resolved, unresolved, and total collisions between dust particles for different Stokes number. The percentage indicates the fraction of resolved collisions.}
    \label{fig:collsion_distribution}
\end{figure*}

Figure \ref{collision_percentage} shows the median dust collision velocity normalised to the sound speed of the gas as a function of the Stokes number. The percentage of resolved pairs is reported in Figure \ref{fig:collsion_distribution}, and depends on the Stokes number, showing a positive correlation as expected. In particular, for St>5 the percentage of resolved pairs is >0.97, while for lower Stokes number the percentage decreases to the point where we don't resolve any pair for the St=0.01 simulation at standard resolution. As expected, the median value of the relative velocity for all pairs is always smaller compared to the resolved ones, as shown in \cite{booth16}. For $\text{St}\geq1$, the median collision
velocities for  all  collisions and that for  the resolved collisions
converge, meaning that the chosen resolution for sampling the dust is enough to ensure that the measured relative velocities accurately represent the relative velocities at zero separation (i.e. the collision velocities). For $\text{St}\leq1$ the median values differ, meaning that higher resolution in sampling the dust component is required. 

\subsubsection{Resolution effects and interpretation of the collision velocity}\label{res_coll}

The collision velocities measured in our simulations are subject to resolution constraints, valid both for gas and dust particles. 

\begin{figure*}
    \centering
    \includegraphics[width=0.485\linewidth]{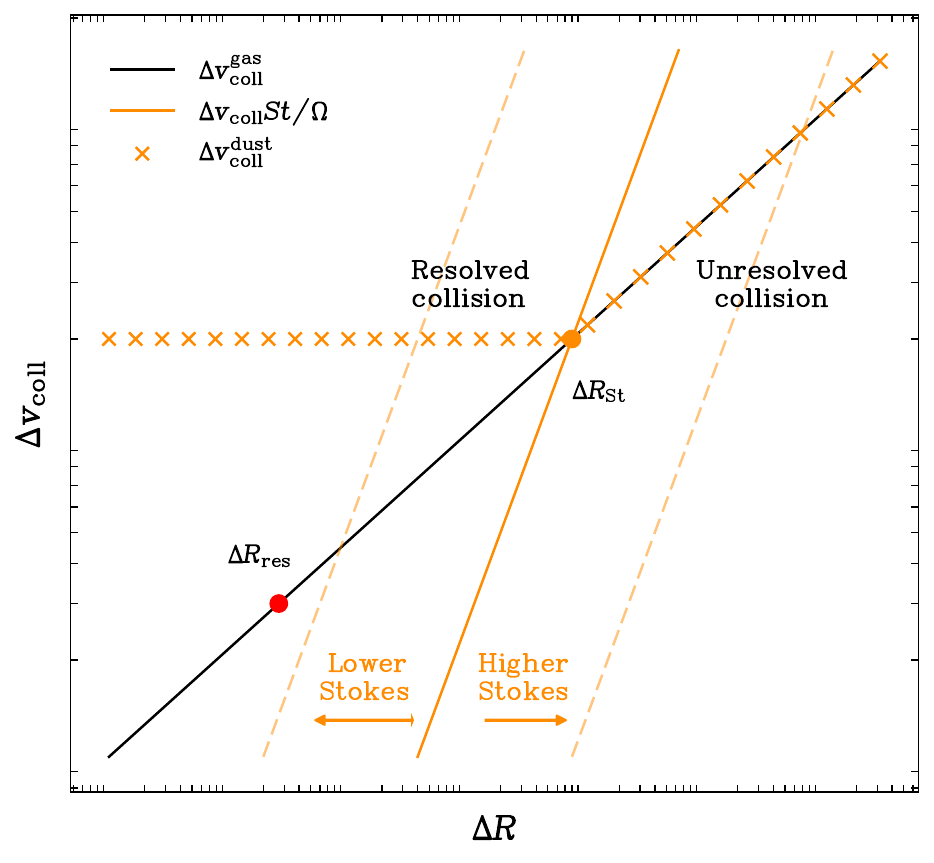}\\
    \includegraphics[width=0.485\linewidth]{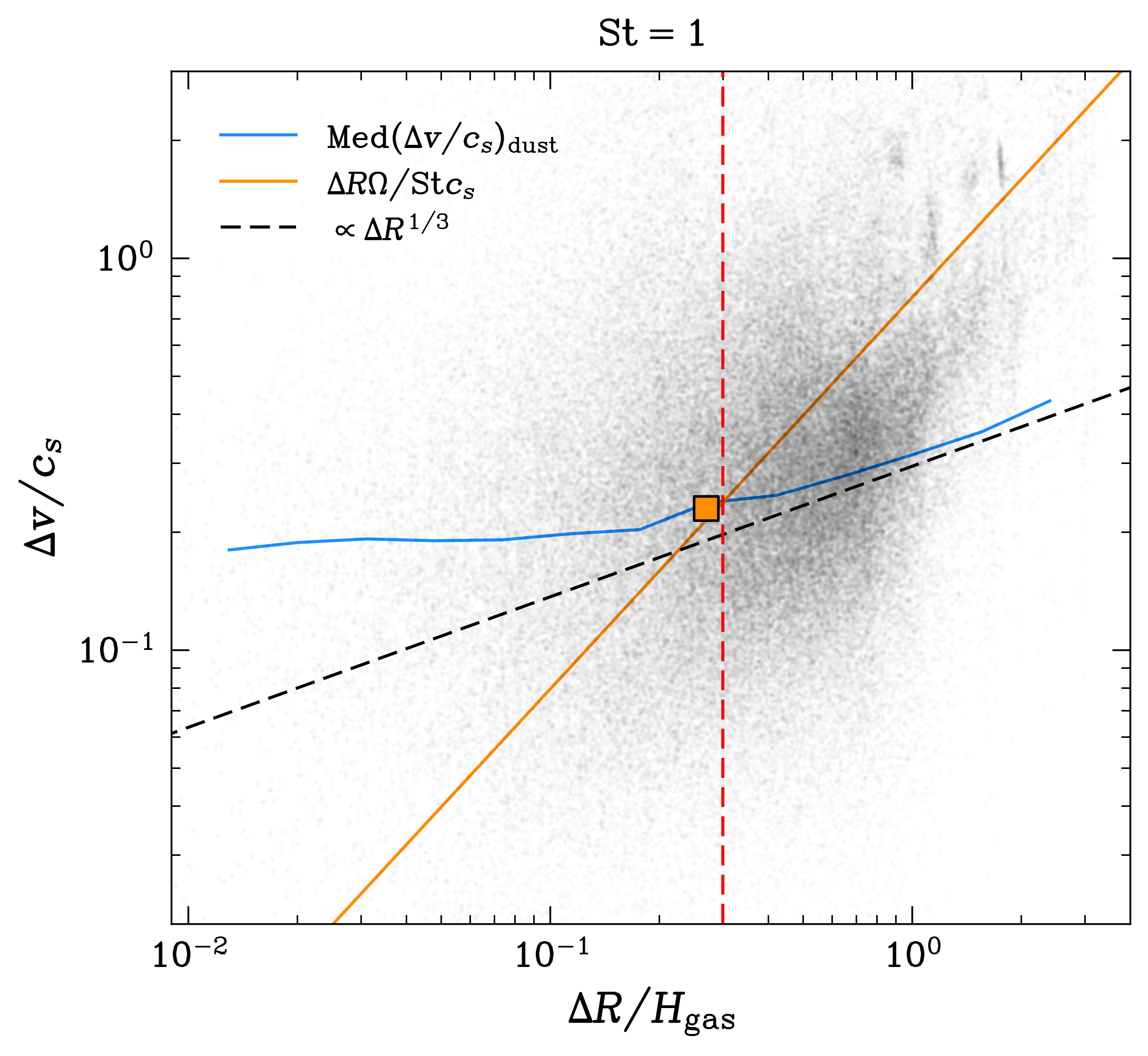}
    \includegraphics[width=0.485\linewidth]{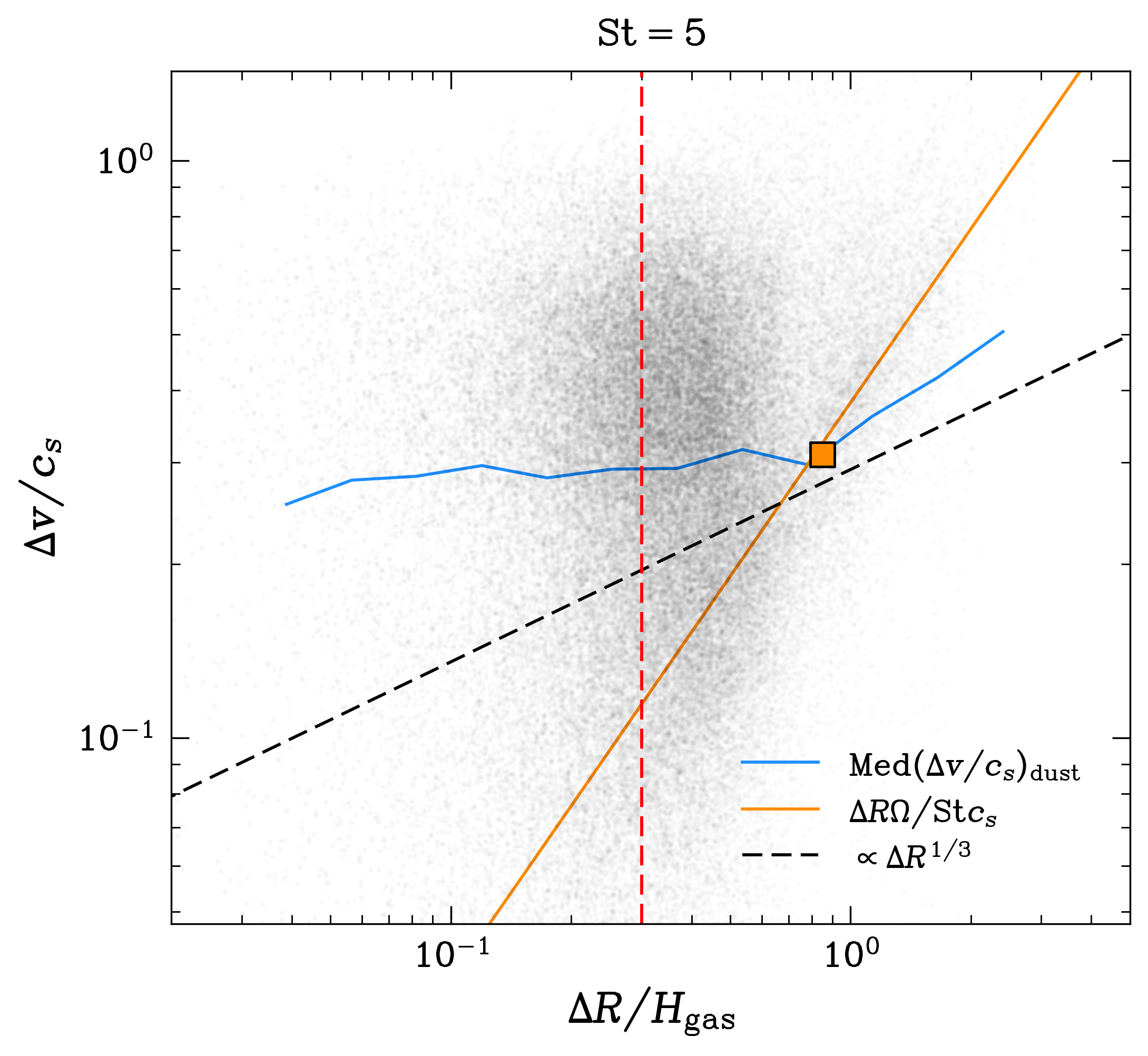}
    \caption{Top panel: Illustrative plot showing gas and dust collisional velocity as a function of particle separation. While gas follows a Kolmogorov scaling $\Delta v\propto \Delta R^{1/3}$ (black line), the collisional velocity of dust particles depends on the Stokes number, and on their separation compared to the stopping length scale $\Delta R_{\rm St}$. Bottom panels: Same illustrative plots but with the data points from the numerical simulations with St$=1$ and St$=5$. The blue line corresponds to the median value of $\Delta v_{\rm coll}^{\rm dust}/c_s$, the orange dot corresponds to $\Delta R_{\rm St}$ and the red vertical line to the $\Delta R_{\rm res}$, as inferred from the power spectrum. }
    \label{fig:sketch}
\end{figure*}

The top panel of figure~\ref{fig:sketch} schematically illustrates how the relative velocity of gas and dust particles depends on particle separation, and how this determines whether a dust collision is numerically resolved. The black line displays the scaling of the gas collisional velocity, which increases with separation following $\Delta v_{\rm gas} \propto \Delta R^{1/3}$, as expected from Kolmogorov turbulence. This scaling reflects the fact that larger eddies induce larger velocity differences between particles separated by larger distances. Dust velocity dispersion should follow that line, down to the scale at which $\Delta R$ is comparable with the stopping distance. 

This condition can be visualised directly in the figure. 
The solid orange line $\Delta v_{\rm coll}\,{\rm St}\,\Omega^{-1}$ 
marks the dividing line between resolved and unresolved collisions: 
pairs with separation $\Delta R$ to the left of its intersection 
with $\Delta v^{\rm gas}$ satisfy $\Delta R < \lambda_{\rm stop}$ 
and are therefore resolved, while those to the right have 
$\Delta R > \lambda_{\rm stop}$ and are unresolved. 
The intersection itself defines the scale $\Delta R_{\rm St}$, 
i.e. the stopping distance associated with that particle pair.
In this regime, the measured dust relative velocity can be interpreted as a reliable estimate of the true collisional velocity\footnote{This is true provided that the measured $\Delta v_{\rm coll}^{\rm dust}$ lies above the gas Kolmogorov scaling $\Delta v^{\rm gas}(\Delta R)$ at the relevant separation. If instead the measured dust velocity falls below the gas line, this would indicate that the gas turbulence driving the relative motions is itself unresolved at that scale, and the inferred collisional velocity would be a lower limit rather than a reliable estimate.}. In this regime, particles are unaffected by drag as they approach each other and therefore this corresponds to the region of the plot where the orange crosses track horizontally to the left on scales less than $\Delta R_{St}$. By contrast, if the particle separation lies to the right of $\Delta R_{\rm St}$, then $\Delta R > \lambda_{\rm stop}$ and the collision is unresolved. In that case, the measured relative velocity might be regarded as an upper limit, since gas drag would damp the particles' relative motion before they physically collide. However, this is not necessarily a strict upper limit: the finite resolution of the simulation means that small-scale motions are not captured, and the true collision velocities could in principle be higher.

The stopping distance depends on the Stokes number. As a result, for small Stokes numbers $\Delta R_{\rm St}$ shifts towards progressively smaller particle separations, making it difficult to measure the collision velocity without decreasing the length scale over which dust relative velocities are sampled. Eventually, the Stokes number becomes sufficiently small that $\Delta R_{\rm St} = \Delta R_{\rm res}$, where $\Delta R_{\rm res}$ denotes the minimum scale at which the simulation is able to resolve turbulent motions in the gas, as inferred from the power spectrum. In this regime, the inferred collisional velocities are no longer physically meaningful, because the turbulent gas motions that drive the particle relative velocities are themselves unresolved by the simulation. It is possible to estimate what is the critical Stokes number at which $\Delta R_{\rm St} = \Delta R_{\rm res}$ for our simulation. Imposing the previous condition, we obtain
\begin{equation}\label{st_crit}
    {\rm St}_{\rm cr} = \frac{\Omega \Delta R_{\rm res}}{\Delta v_{\rm coll}}.
\end{equation}
Figure \ref{pspectrum} shows that $\Delta R_{\rm res}=H/3$, hence
\begin{equation}
    {\rm St}_{\rm cr} = \frac{c_s}{3\,\Delta v_{\rm coll}}.
\end{equation}
This shows that the critical Stokes number depends on the characteristic collisional velocity adopted, and therefore should be regarded as an order-of-magnitude estimate rather than a sharp threshold. For a typical value of  collisional velocity $\Delta v_{\rm coll}=0.2c_s$ (see Fig.\ref{fig:cdcs_median}), the critical Stokes number is of the order of 1.6. 

One might think that an additional resolution requirement is that the separation between two dust particles should be larger than  $\Delta R_{\rm res}$. However, this condition is not necessary as long as the collision itself is resolved according to the stopping-distance criterion discussed above. This can be understood from the behaviour of the collisional velocity as a function of particle separation. In the regime where the curve becomes flat, the measured $\Delta v_{\rm coll}$ has already converged to the physical value set by the particle dynamics driven by larger scales. Turbulent motions at scales smaller than $\Delta R_{\rm res}$ would induce velocity fluctuations that follow the Kolmogorov scaling and therefore correspond to progressively smaller velocity differences. As a result, unresolved turbulent eddies would excite relative velocities that are smaller than the collisional velocity already measured. Consequently, the lack of resolution at scales below $\Delta R_{\rm res}$ does not bias the inferred collision velocity in this regime. Once the collisional-velocity curve has reached a plateau, the measured value represents the true physical collision velocity, even if the smallest turbulent scales in the gas are not fully resolved.

The bottom panels of Fig.~\ref{fig:sketch} show the same illustrative plot, now including the data points from the simulations. In particular, we show examples for particles with ${\rm St}=1$ and ${\rm St}=5$. For the ${\rm St}=1$ case, the median collisional velocity of dust particles (blue line) follows the Kolmogorov scaling (black dashed line) for $\Delta R > \Delta R_{\rm St}$, and flattens at smaller separations. In this case, as expected from eq.~\eqref{st_crit}, $\Delta R_{\rm St} \simeq \Delta R_{\rm res}$. The ${\rm St}=5$ case instead provides an ideal example of the resolution requirements discussed above. The Kolmogorov scaling is recovered at large separations, after which the dust collisional velocity flattens for $\Delta R < \Delta R_{\rm St}$. Importantly, the curve remains flat even when crossing $\Delta R_{\rm res}$. This occurs because, at these Stokes numbers, dust particles are largely decoupled from the gas, so that unresolved turbulent motions at scales smaller than $\Delta R_{\rm res}$ do not affect the measured collisional velocity.

\subsubsection{2D vs 3D}

Our results can be compared with the two-dimensional SPH simulations of \citet{booth16}, who studied dust collisional velocities in gravitationally unstable discs using a similar physical framework but a different numerical setup. Despite the difference in dimensionality and code, the agreement between the two studies is very good.

In the regime where dust--dust collisions are numerically well resolved, namely for $\mathrm{St}>1$, we find an excellent agreement with \citet{booth16}. Both studies show that the median collisional velocity increases with Stokes number and converges once the stopping distance exceeds the resolution scale for the gas, $\Delta R_{\rm res}$. This agreement is particularly reassuring given that the simulations are performed with different codes and in different dimensionality (2D versus 3D), indicating that collisional velocities in this regime are robust.

For $\mathrm{St}\le1$ the values diverge as the resolution scale becomes greater than the stopping length; since $\Delta R_{\rm res}$ is around the smoothing length and thus comparable to the length scale on which pairs of dust particles are sampled, in this regime the bulk of collisions are unresolved in our simulations. Indeed our global 3D simulations resolve a somewhat smaller fraction of particle pairs than the 2D simulations of \citet{booth16}, as a consequence of the lower numerical resolution per unit volume in three dimensions. In particular, the 2D simulations resolve approximately $30\%$ of collisions at $\mathrm{St}=0.1$, $50\%$ at $\mathrm{St}=0.3$, and $70\%$ at $\mathrm{St}=1$. In contrast, in our 3D simulations only a few per cent of collisions are resolved for $\mathrm{St}\le0.3$, increasing to $\sim20$--$25\%$ at $\mathrm{St}=1$ (Fig. \ref{fig:collsion_distribution}).

In \citet{booth16}, the critical Stokes number $\mathrm{St}_{\rm cr}$ can be estimated from the average $(h/H)_{\rm gas}$, which is reported to be $0.2$ for the standard resolution case ($N_g = 10^6$) and $0.08$ for the high-resolution case ($N_g = 16 \times 10^6$). These values correspond to a critical Stokes number of $1$ and $0.45$, respectively. This comparison is particularly significant because it demonstrates that, in both 2D and 3D, the simulation results follow the same $\sqrt{\mathrm{St}}$ scaling  for all cases where $\mathrm{St} > \mathrm{St}_{\rm cr}$. The fact that this scaling is consistently recovered once the stopping distance exceeds the resolution scale, confirms that the collisional velocities in this regime are physically robust and independent of the numerical setup or dimensionality.

\subsection{Dust velocity dispersion}
The dust velocity dispersion is a key quantity for assessing the stability of the dust layer against gravitational collapse. In the context of two-fluid gravitational instability, it plays the role of an effective sound speed, providing pressure-like support against self-gravity. However, it is important to note that the velocity dispersion is not a single-valued quantity: it depends on the spatial scale $\Delta R_{\rm prob}$ over which it is measured, since larger scales are increasingly contaminated by bulk motions such as shear and systematic flows within the spiral arms.

The dependence on Stokes number at a fixed probing scale $\Delta R_{\rm prob}$ is governed by the ratio between $\Delta R_{\rm prob}$ and the stopping length $\Delta R_{\rm St}$. When $\Delta R_{\rm St} > \Delta R_{\rm prob}$, the dust is coupled to the gas down to a minimum scale $\Delta R_{\rm St}$ that exceeds the probing scale; the dust therefore acquires a velocity dispersion characteristic of this larger decoupling scale, imprinting the $\propto \sqrt{\rm St}$ scaling discussed in Section~\ref{res_coll}. As the Stokes number decreases, $\Delta R_{\rm St}$ approaches $\Delta R_{\rm prob}$, and below this critical value — where $\Delta R_{\rm St} < \Delta R_{\rm prob}$ — the velocity dispersion reaches a plateau. In this regime, the dominant contribution to the dust velocity dispersion comes from eddies whose turnover scale matches the probing scale, and since this scale is fixed, the velocity dispersion becomes independent of the Stokes number.

The choice of probing scale therefore determines where the plateau occurs. Selecting the physically relevant scale is non-trivial: for the purpose of assessing the onset of dust-driven gravitational instability, a natural choice is the dust Jeans length $\lambda_{\rm J,d}$, as this is the scale on which self-gravity and pressure-like support compete. However, in this work we evaluate the dust velocity dispersion on a particle-by-particle basis, using the dust smoothing length $h_d$ as the probing scale, as done in \citet{longarini23b} and \citet{rowther24}. 

Nevertheless, we caution that the exact position of the plateau in Stokes number is sensitive to the probing scale and, consequently, to the numerical resolution. Since $h_d$ is set by the number of dust particles and their local density, a lower resolution simulation will have a larger $h_d$, shifting the transition between the $\sqrt{\rm St}$ regime and the plateau to higher Stokes numbers. Therefore, the plateau we observe may be displaced towards lower Stokes numbers in higher-resolution simulations, and the results in this regime should be interpreted as 
resolution-dependent rather than as a robust physical 
prediction.

We compute the dust velocity dispersion $c_d$ following
\begin{equation}\label{dustdisp_SPH}
    c_{d,i}^2 = \sum_{j=1}^{N_{\rm neigh}} m_j 
    \frac{(v_{d,i} - v_{d,j})^2}{\rho_j} 
    W_{ij}(h_i),
\end{equation}
as done in \citet{longarini23b} and \citet{rowther24}. This quantity is computed within the SPH framework using the kernel $W_{ij}$, and represents the root-mean-square of the relative velocity of dust particle $i$ with respect to its neighbours, weighted by the local density and kernel function. 

\begin{figure}
    \centering
    \includegraphics[width=\linewidth]{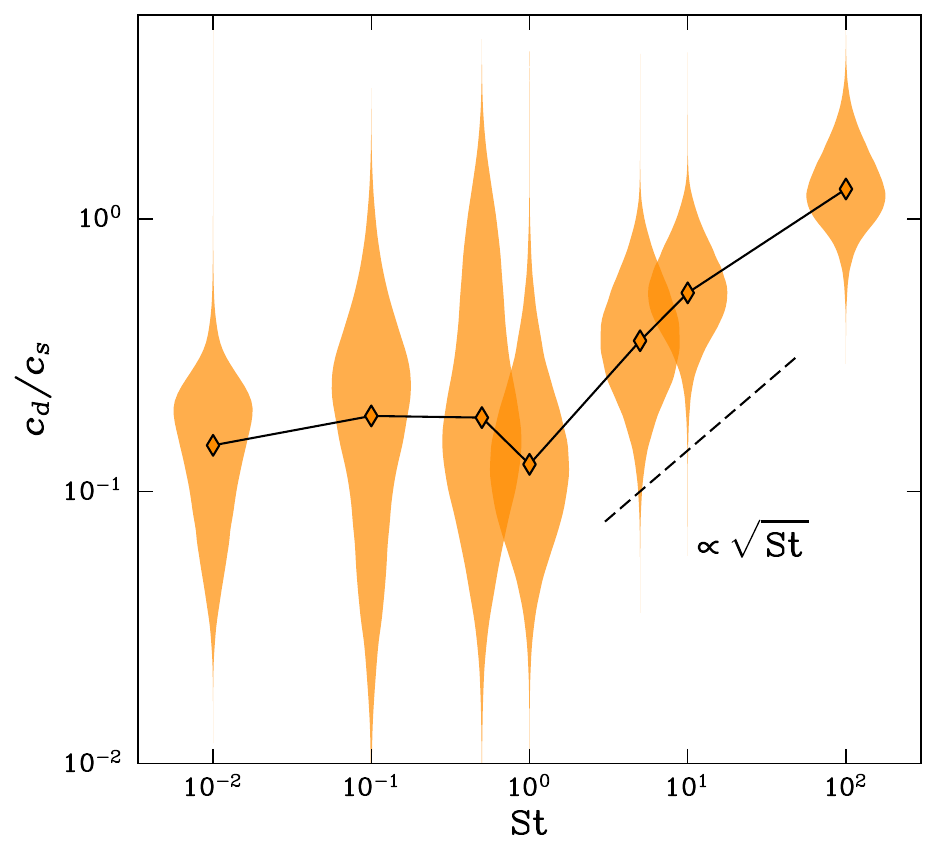}
    \caption{Violin plots of the distributions of dust velocity dispersion normalised to the gas sound speed, $c_d/c_s$, for different Stokes numbers. The diamonds correspond to the median values of the distributions, showing a minimum around $\mathrm{St} \simeq 1$ and a plateau for $\mathrm{St} < 1$. For $\mathrm{St} \gtrsim 1$, the median follows the expected $\propto \sqrt{\mathrm{St}}$ scaling (see Section~\ref{res_coll}). The plateau at low Stokes numbers is a resolution-dependent feature: it arises because the velocity dispersion is measured on the dust smoothing length $h_d$, which sets a floor on the probing scale. Higher-resolution simulations, with smaller $h_d$, would shift the onset of the plateau to lower Stokes numbers, and the values of $c_d/c_s$ in the plateau regime should therefore be interpreted as upper limits.}
    \label{fig:cdcs_median}
\end{figure}

Figure~\ref{fig:cdcs_median} shows the distribution of the dust velocity dispersion normalised to the gas sound speed, $c_d/c_s$, as a function of the Stokes number, together with the median values (diamonds). The trend is consistent with the picture outlined above. As discussed above, the exact position of the transition between the $\sqrt{\mathrm{St}}$ regime and the plateau is resolution dependent, and higher-resolution simulations would be expected to shift it towards lower Stokes numbers.

\subsubsection{Anisotropy of the velocity dispersion}
To thoroughly understand the dust velocity dispersion, we decompose it into the three directions, radial, azimuthal and vertical. We compute these quantities using eq. \eqref{dustdisp_SPH} using only one component of the dust particles' velocity.

\begin{figure*}
    \centering
    \includegraphics[width=0.475\linewidth]{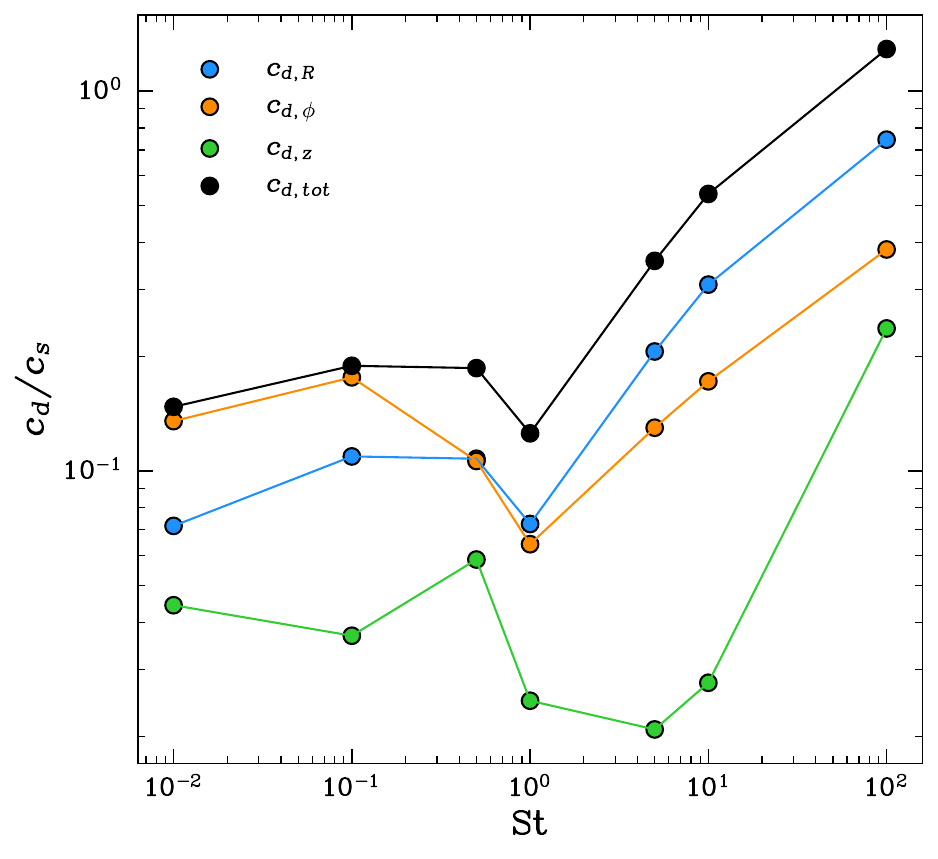}
    \includegraphics[width=0.475\linewidth]{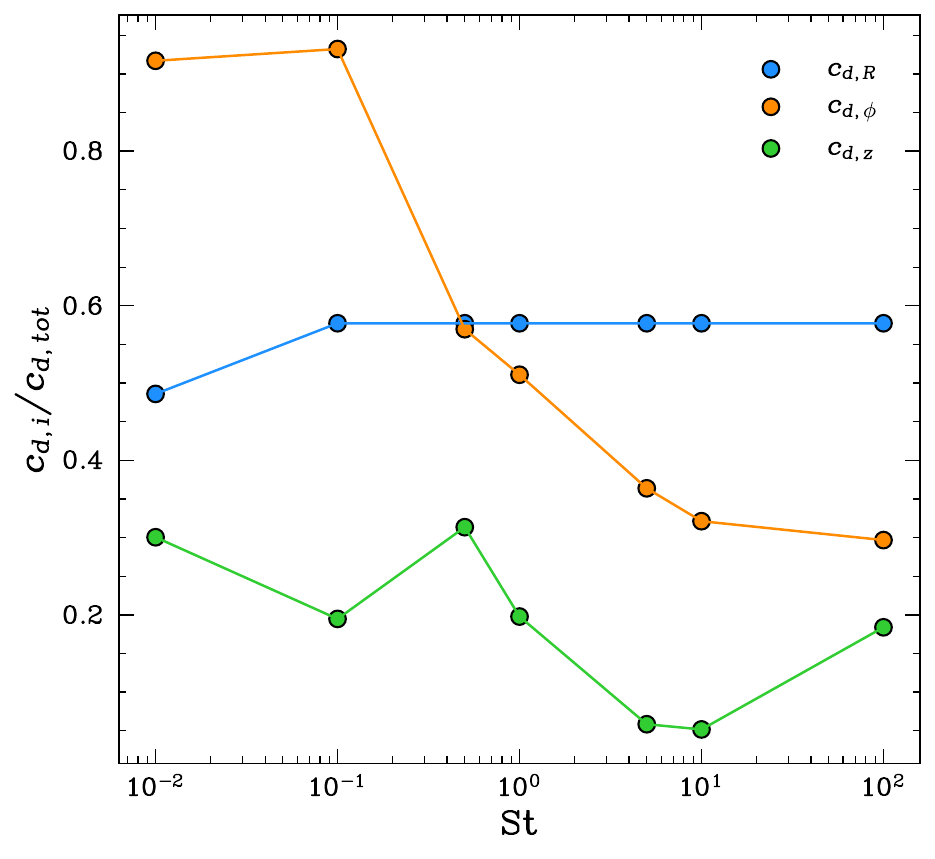}
    \caption{Dust dispersion velocity in the three directions (radial, azimuthal and vertical) as a function of Stokes number.}
    \label{fig:disp_components}
\end{figure*}

The left panel of figure \ref{fig:disp_components} shows the median value of the velocity dispersion in the three different directions compared to the total one as a function of the Stokes number. The plot consistently shows that the velocity dispersion is mainly driven by the azimuthal and radial components, while the vertical one is always smaller. The higher azimuthal dispersion for low Stokes number can be driven by Keplerian shear, as it decreases with Stokes number. The right panel of figure \ref{fig:disp_components} shows the ratio of the velocity dispersion in the three directions with respect to the total one, as a function of the Stokes number. We recall that the velocity dispersion is normalised so that $c_{d,R}^2+c_{d,\phi}^2+c_{d,z}^2 = c_d^2$. The figure shows that for small Stokes number St $<1$, the velocity dispersion is dominated by radial and azimuthal motions, with similar magnitude. In particular, we note that for St $\leq 0.1$ the azimuthal component is slightly bigger than the radial one. Conversely, for St $>1$ the radial component dominates, because the drag force is negligible and the interaction between spiral arms and dust particles is just gravitational. 

As just shown, the vertical velocity dispersion is negligible compared to the other components. This is particularly interesting, showing that gravitational instability is predominantly planar. We note, however, that 3D GI spiral waves are known to excite coherent vertical circulations in the form of poloidal rolls at $|z| < H$ \citep{riols18}, which could in principle contribute to the vertical velocity dispersion. The fact that this contribution appears negligible in our simulations suggests that these motions do not significantly affect the dust dynamics, and explains why the 3D results match well with the 2D ones of \citet{booth16}.

\section{Discussion}\label{sec_discussion}

\subsection{Dust growth in GI spirals}
The collision velocity between dust particles plays a crucial role in determining whether collisions lead to sticking, bouncing, or fragmentation. If the relative velocity is sufficiently high, collisions result in fragmentation into smaller grains, while at moderate velocities they may instead lead to bouncing or growth. Accurately determining the fragmentation threshold is non-trivial, as it depends on the intrinsic properties of the grains, such as size, composition, and porosity. In this work, we do not attempt to model these microphysical properties in detail and therefore adopt representative fragmentation velocities of $v_{\rm frag}\simeq1,3,10\mathrm{ms^{-1}}$ \citep{birnstiel24}.

The collisional velocities measured in our simulations, even in the most favourable cases, are typically too high to allow sustained dust growth. The primary reason for this is numerical. As shown in Fig.~\ref{fig:collsion_distribution}, only a small fraction of particle pairs satisfy the resolution criterion required to reliably measure collision velocities, and these correspond to high-velocity encounters. As a result, our measurements are biased towards larger collisional velocities. 

We have no reason to expect dust collisional velocities to deviate from the $\mathrm{St}^{1/2}$ scaling at low Stokes number, as discussed in section \ref{sec:pspec}. This scaling is very well recovered for high Stokes number, and there is an excellent agreement between 2D and 3D simulations. Since the scaling holds, we extrapolate the value of collisional velocity at low Stokes number, as shown in Fig.~\ref{fig:collsion_distribution}. We find that for $\mathrm{St} \sim 0.3$, $\Delta v / c_s \sim 0.1$, and assuming a typical value of $c_s \sim 100\,\mathrm{m\,s^{-1}}$, this corresponds to $\Delta v \sim 10\,\mathrm{m\,s^{-1}}$, that is the assumed fragmentation threshold. We underline that these are order-of-magnitude estimates, as they rely on an extrapolation of the $\mathrm{St}^{1/2}$ scaling and on a simplified choice of the sound speed. 
Nevertheless, this result shows that it is in principle possible for dust particles to grow up to $\mathrm{St} \sim 0.1$ in self-gravitating discs, since  collisional velocities remain close to or below the fragmentation threshold, as already pointed out by \citet{booth19}.

A second important limitation for assessing dust growth is that our simulations consider a monodisperse dust population, i.e. collisions are evaluated only between particles with the same Stokes number. However, a realistic dust population is intrinsically multidisperse, and relative velocities between grains of different sizes can significantly modify collisional outcomes. Addressing this problem robustly requires local, high-resolution simulations that can accurately resolve dust–dust collisions, combined with a multidisperse dust model \citep{Leedham26prep}.

\subsection{Dust collapse in GI spirals}
Planetesimals and planetary cores formation through dust collapse within gas spiral arms has been proposed as a viable early planet formation scenario \citep{rice04, longarini23b,rowther24}. Linear stability analysis of a two-phase flow \citep{longarini23a} suggests that the second cold component, dust in this case, is gravitationally unstable when it is cold and abundant enough relative to the dominant component (the gas).

\begin{figure*}
    \centering
    \includegraphics[width=0.33\linewidth]{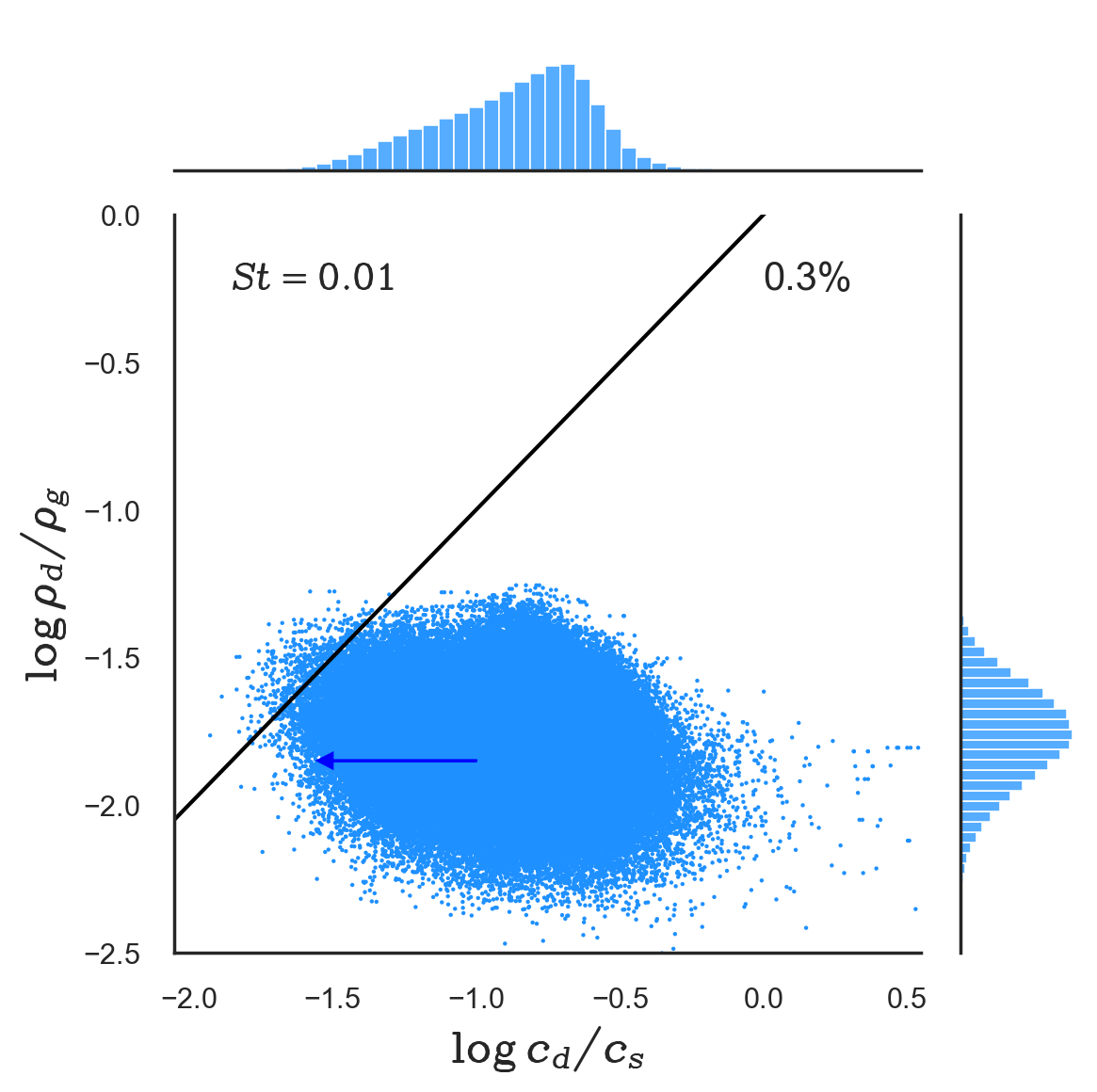}
    \includegraphics[width=0.33\linewidth]{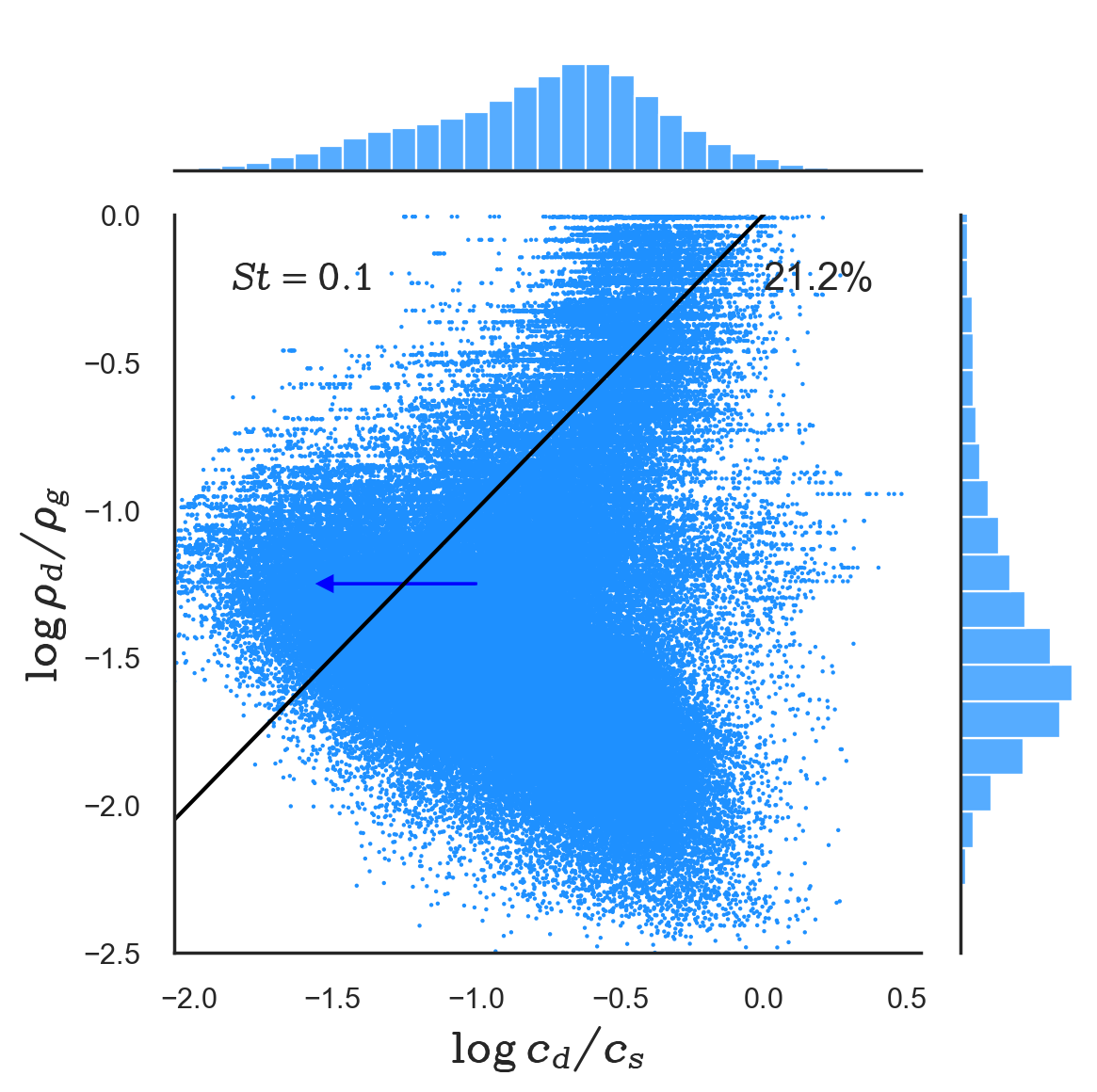}
    \includegraphics[width=0.33\linewidth]{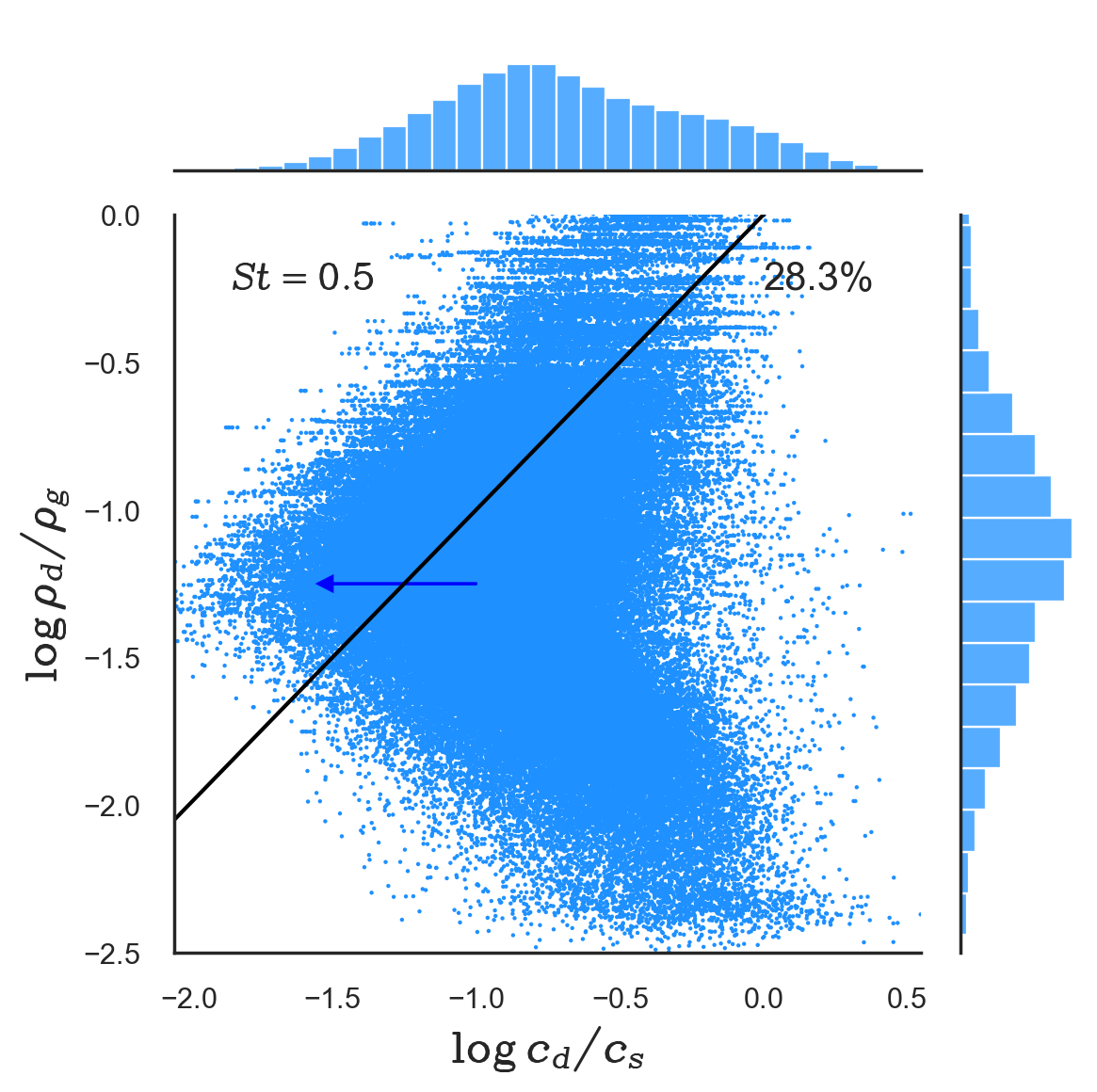}
    \includegraphics[width=0.33\linewidth]{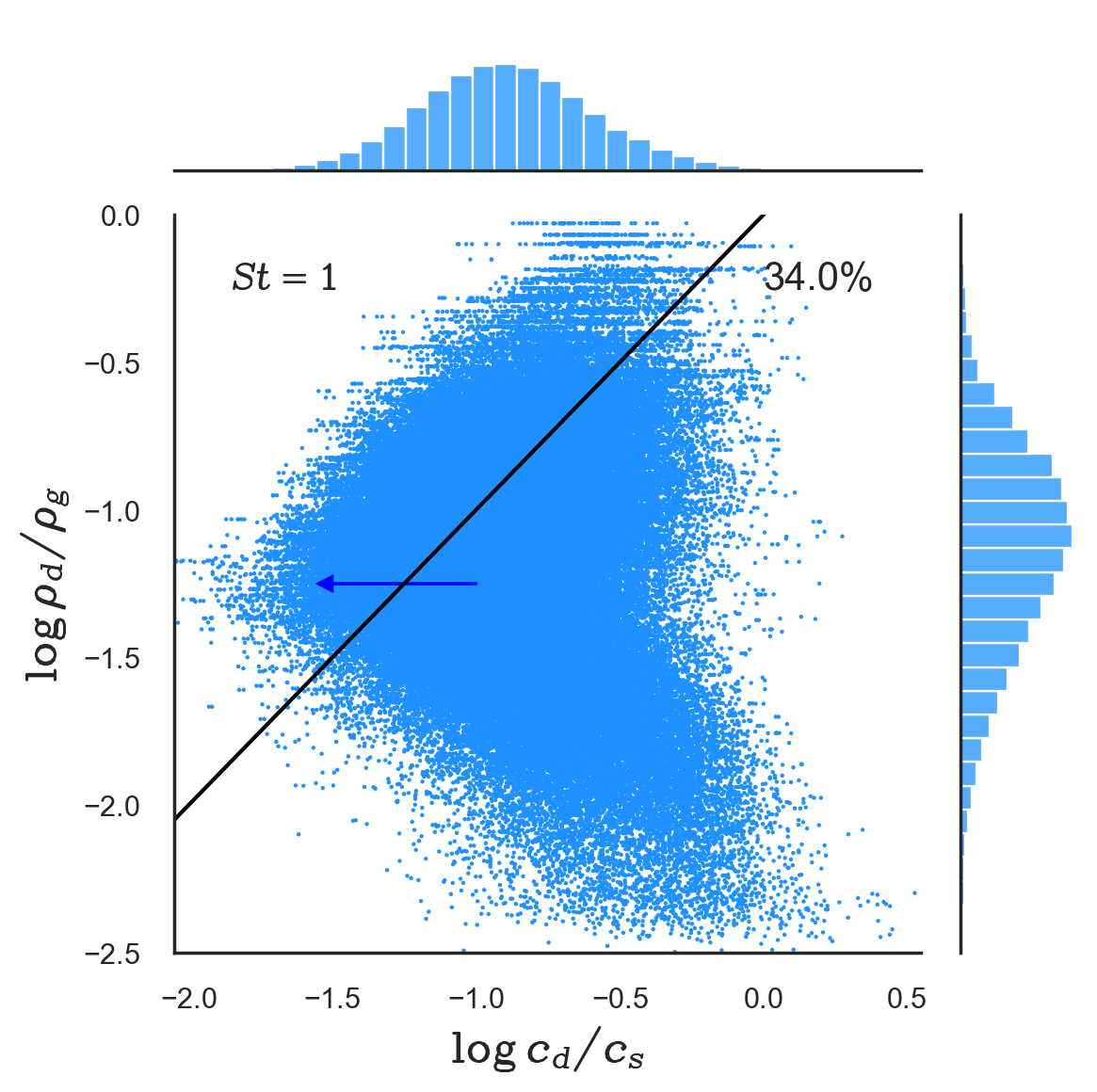}
    \includegraphics[width=0.33\linewidth]{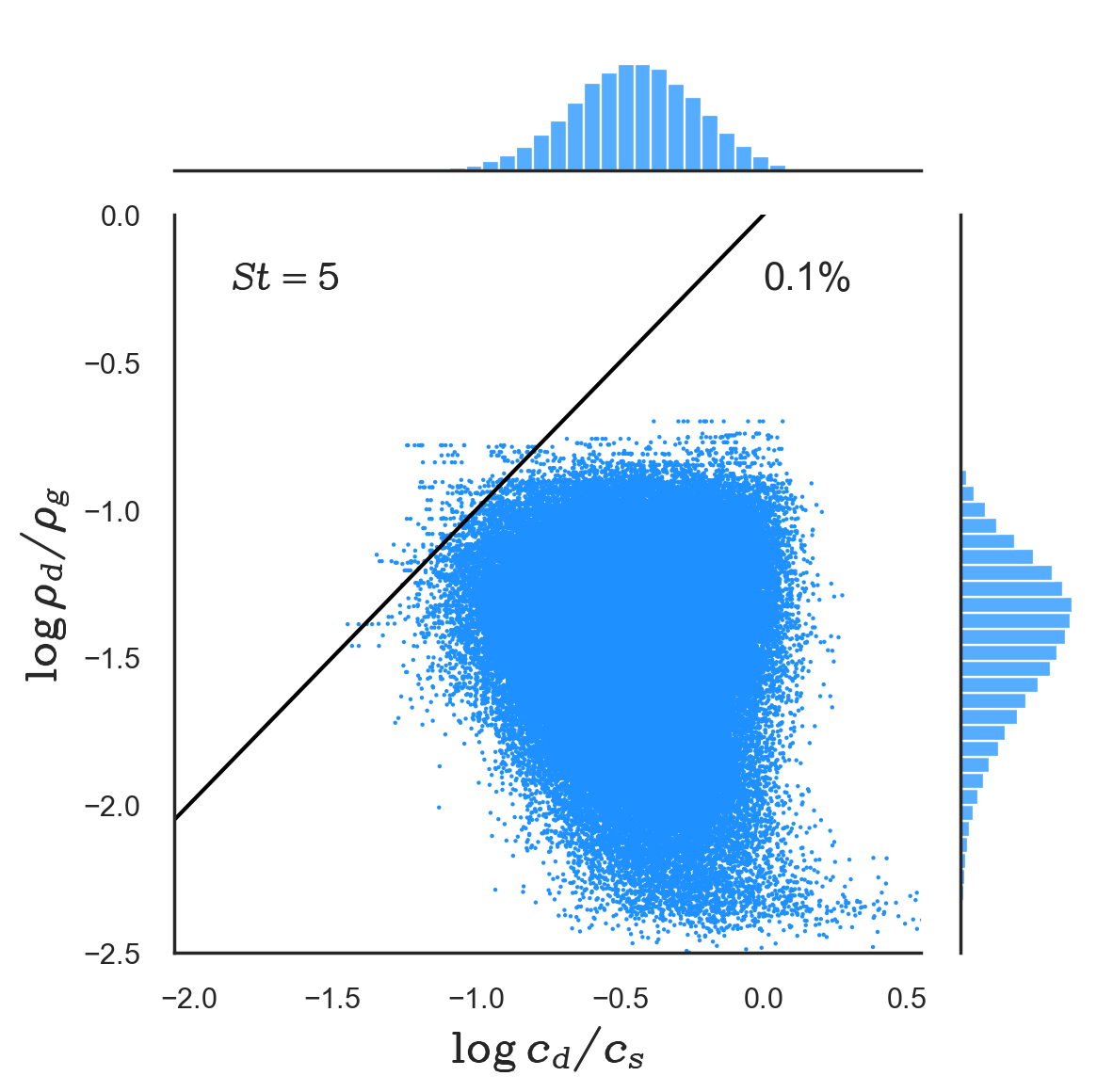}
    \includegraphics[width=0.33\linewidth]{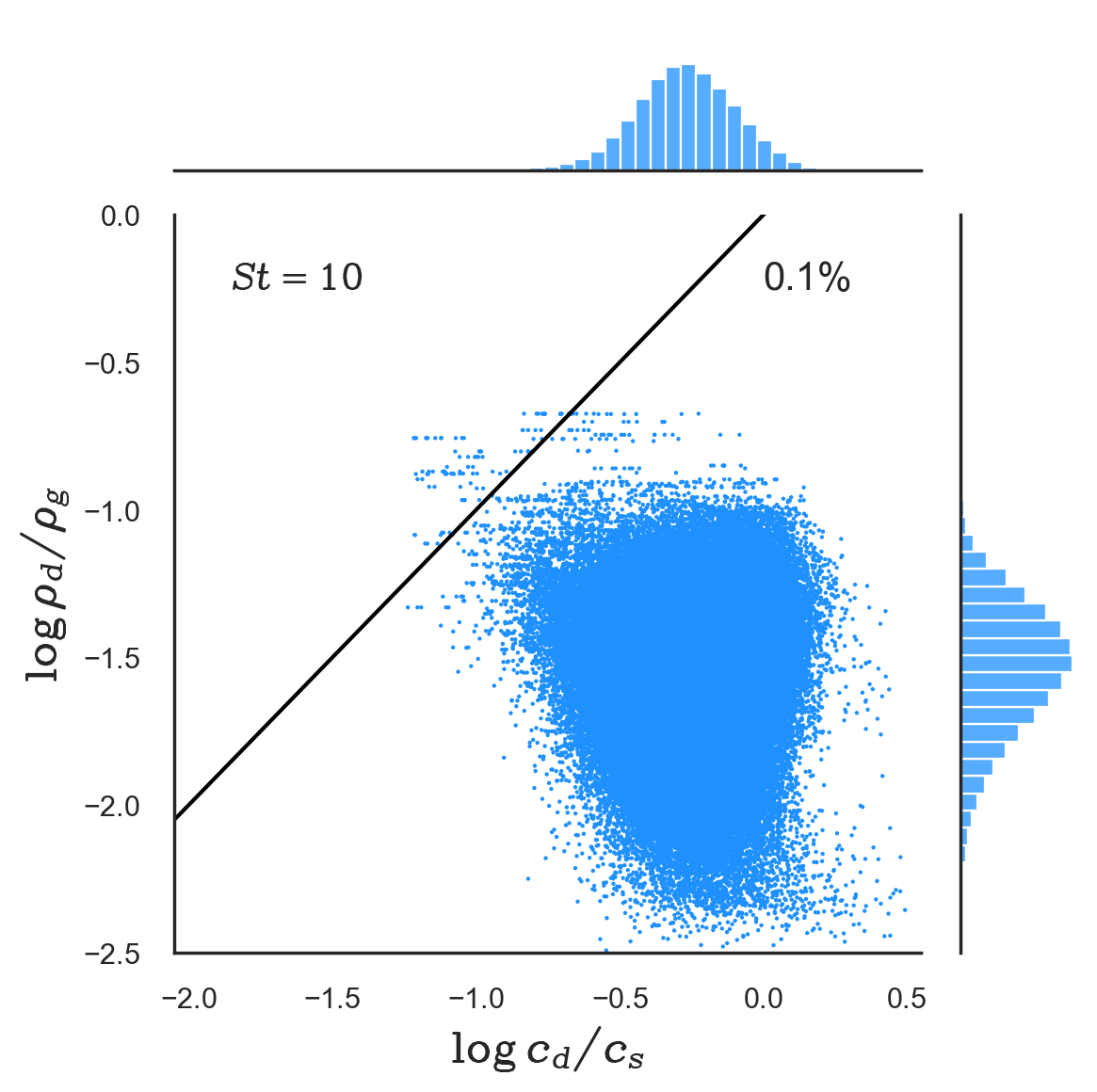}
    \includegraphics[width=0.33\linewidth]{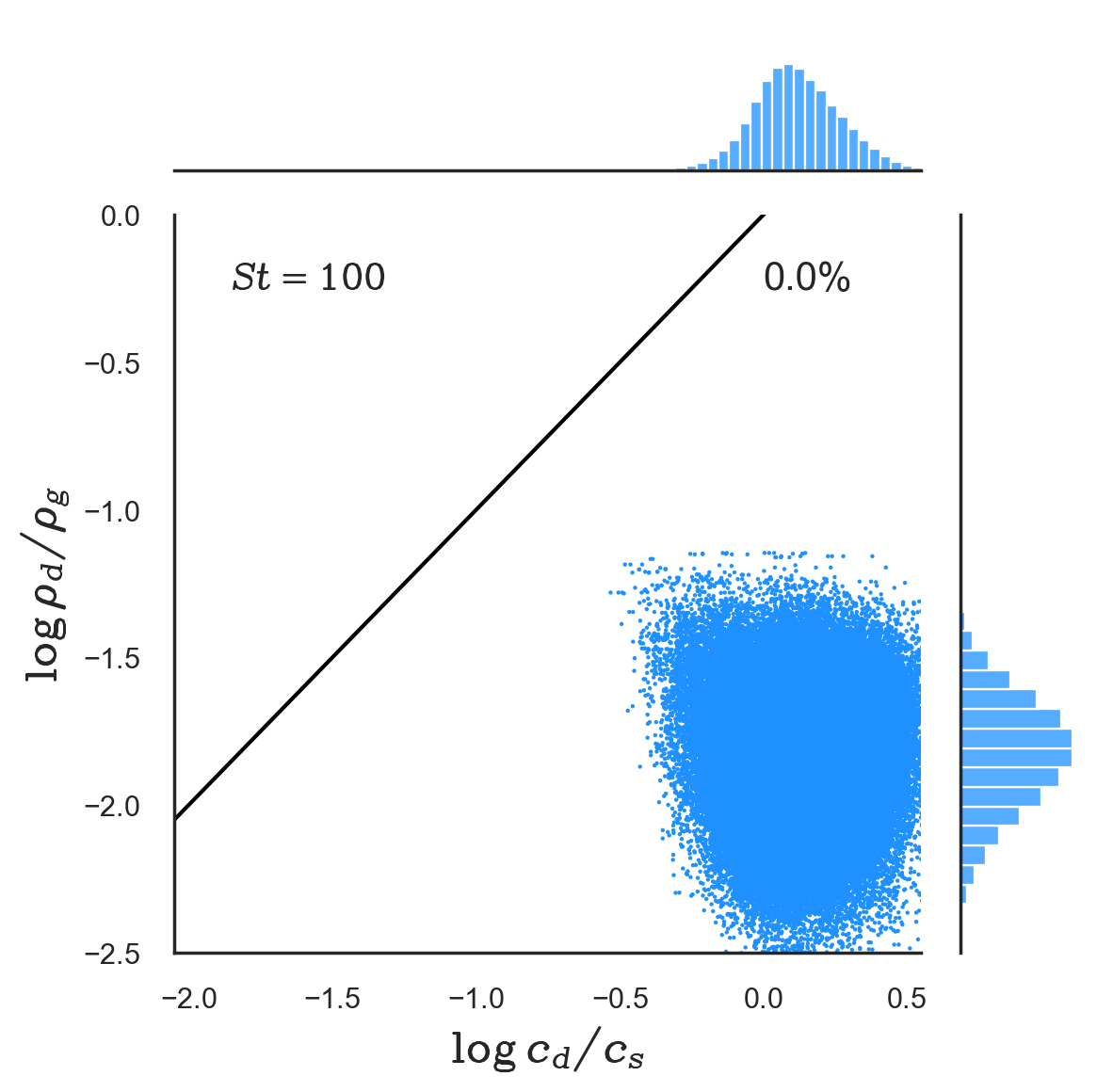}
    
    \caption{Dust particles in the simulations shown in the dust-to-gas ratio--velocity dispersion plane. The black line marks the threshold between gas-driven and dust-driven gravitational instability (below and above the line, respectively). For $\mathrm{St} \geq 1$, the velocity dispersion is reliably measured, and the fraction of particles above the instability threshold can be considered robust. For $\mathrm{St} \lesssim  1$, however, the measured velocity dispersion is set by the turbulent gas velocity at the fixed probing scale $h_d$ rather than at the physical decoupling scale, and should therefore be regarded as an upper limit (see blue lines). Consequently, the fraction of unstable particles in this regime represents a lower limit on the true unstable fraction: higher-resolution simulations, which would probe smaller scales, may find a larger fraction of particles satisfying the instability criterion.}
    \label{fig:dustinst}
\end{figure*}

\begin{figure*}
    \centering
    \includegraphics[width=0.75\linewidth]{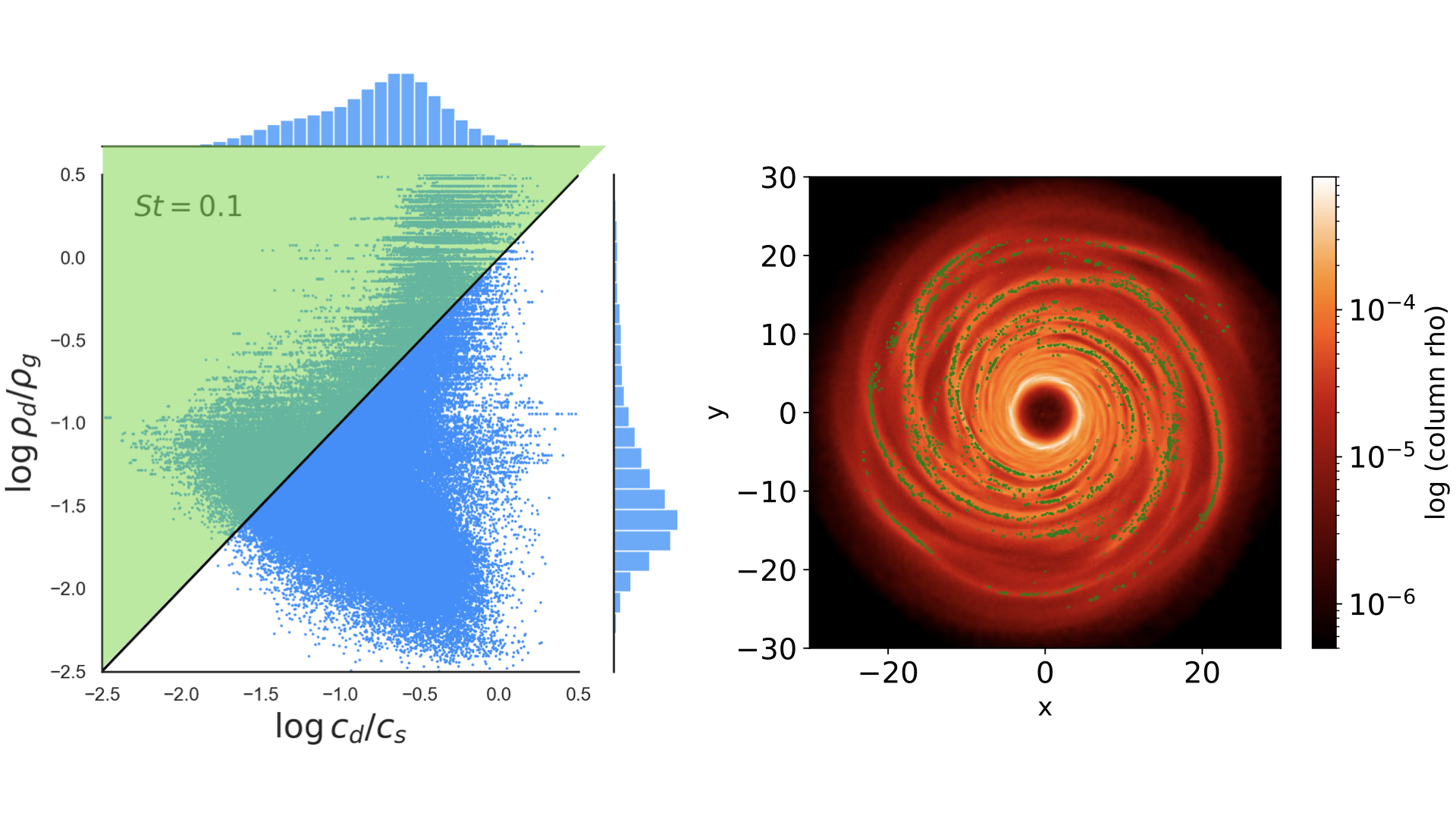}
    \includegraphics[width=0.75\linewidth]{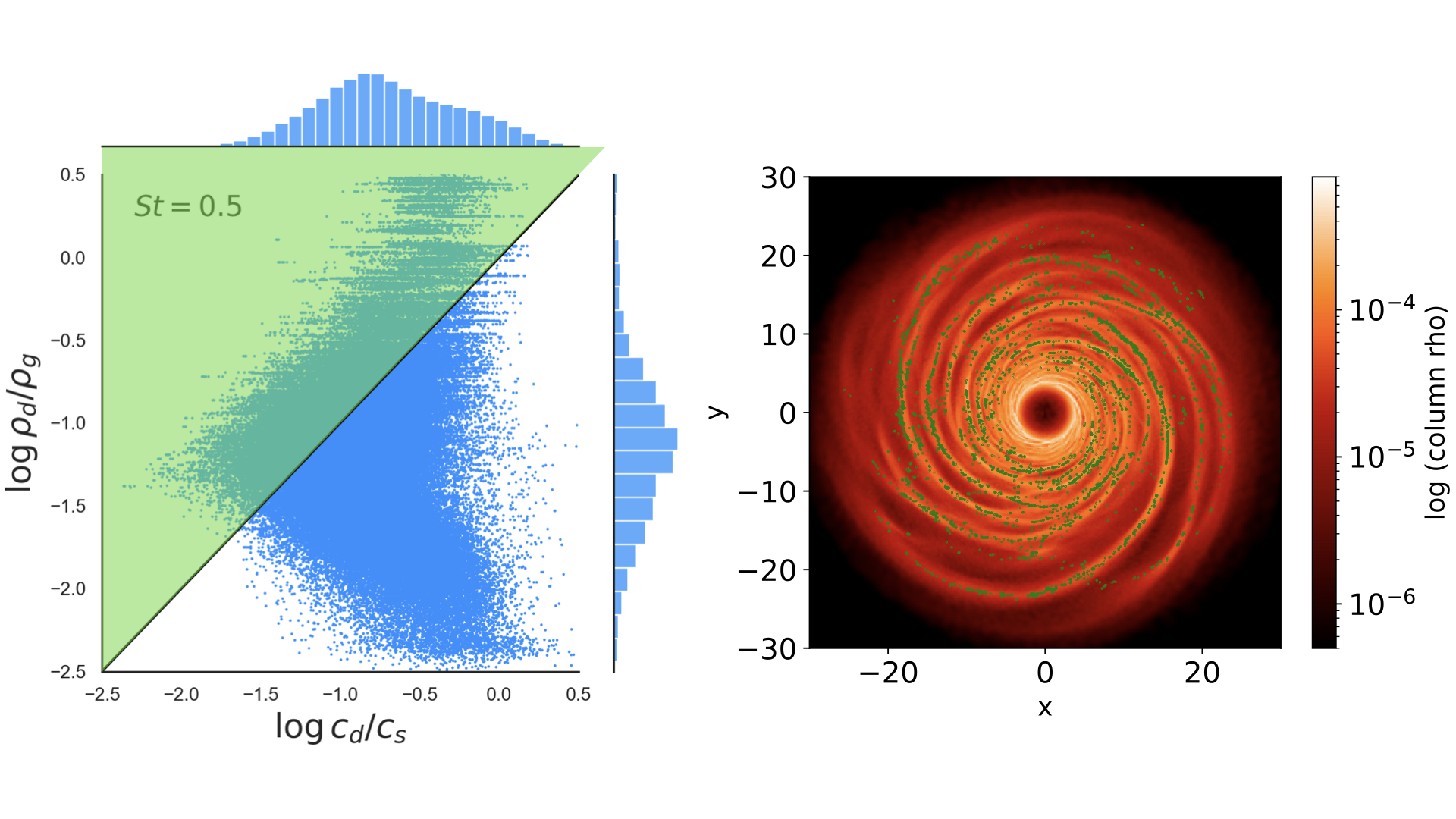}
    \includegraphics[width=0.75\linewidth]{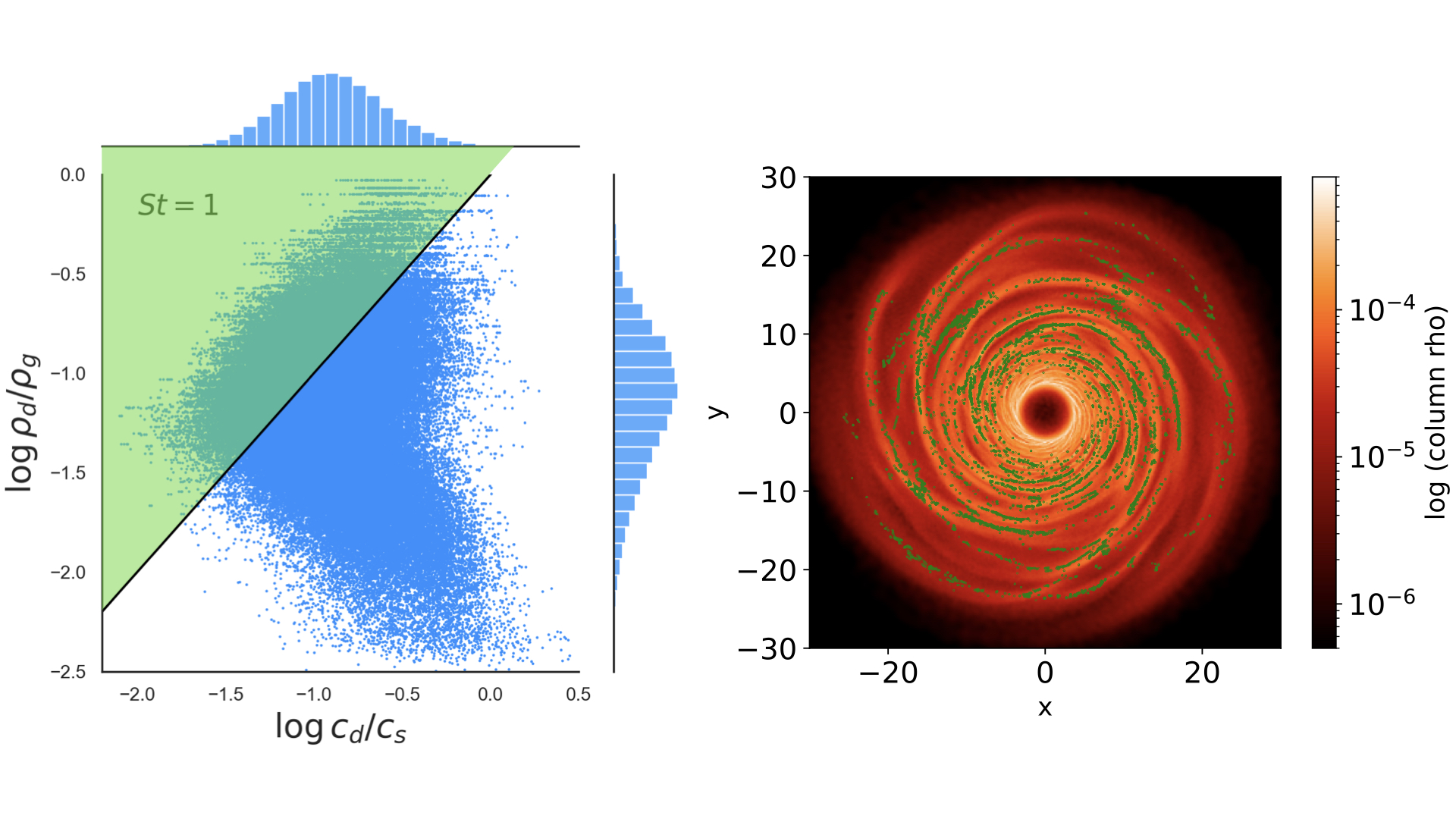}
    \caption{Spatial distribution of dust particles that are linearly unstable to the dust-driven gravitational instability for $\mathrm{St}=0.1$ (top), $\mathrm{St}=0.5$ (middle) and $\mathrm{St}=1$ (bottom). In the left panels, dust particles are shown in the $\rho_{\mathrm{d}}/\rho_{\mathrm{g}}$--$c_{\mathrm{d}}/c_{\mathrm{s}}$ plane; particles classified as unstable according to the dust-driven GI criterion are highlighted in green. The same particles are highlighted in green in the right panels, overlaid on the gas surface density.
}
    \label{fig:zoomin_dustspirals}
\end{figure*}

In particular, assuming that the disc is marginally gravitationally unstable ($Q \sim 1$), the instability criterion reads
\begin{equation}
    \frac{\rho_d}{\rho_g} > \frac{c_d}{c_s},
\end{equation}
where the  scaling arises because, at $Q\sim1$, the Jeans 
length of the gas is comparable to the disc scale height, so that the same condition applies to both phases simultaneously. 
Physically, a smaller sound speed $c_s$ requires a smaller dust-to-gas ratio to trigger collapse, while a larger dust velocity dispersion $c_d$ stabilises the dust layer against its own self-gravity. Figure \ref{fig:dustinst} shows the position of dust particles of the 7 simulations in the $\rho_d/\rho_g-c_d/c_s$ space. The black line corresponds to the dust instability threshold: particles above that line are linearly unstable. The fraction of those particles is written in the plot. As expected, the fraction of particles above the black line is negligible for St$=0.01$ and St$\geq 5$. In the intermediate regime, there is a significant fraction of particles in the dust-driven GI regime, meaning that dust itself is gravitationally unstable. This is the best regime for dust collapse to happen. In figure \ref{fig:zoomin_dustspirals} we map where the particles in the dust-driven GI regime are spatially located in the simulation. As expected, the particles are located within gas spiral arms. 

We therefore conclude that dust particles in the intermediate stokes regime St $\in (0.1,1)$ are likely to undergo collapse, forming planetary cores with masses $M_\text{core}\sim1-10\text{M}_\oplus$, as predicted from linear theory \citep{longarini23a}, and verified in numerical simulations \citep{longarini23b,rowther24,baehr22,rice25}.

\subsection{Summary: planet formation through dust collapse}

To assess the conditions for dust to undergo gravitational collapse inside GI spiral arms, it is useful to summarise the dust--gas dynamics as a function of Stokes number. Across the different coupling regimes, the outcome is determined by the competition between (i) dust trapping in spiral arms, which enhances $\rho_d/\rho_g$, and (ii) dust excitation, which increases the dust velocity dispersion $c_d$ (and thus stabilises the dust layer against collapse).

\begin{itemize}
    \item \textbf{$\mathrm{St}<0.1$ (tight coupling):}
    Dust grains are strongly coupled to the gas and largely trace the gas flow. As a result, the dust-to-gas ratio remains close to the global value, with only modest local enhancements, as efficient trapping requires a stopping time comparable to the dynamical time. In this regime, the dust velocity dispersion is set by the gas turbulent floor at the scale where the two phases are coupled. For these small Stokes numbers, the particles are advected coherently by all resolved turbulent eddies, and their relative motions are effectively damped by drag. Consequently, the dust velocity dispersion reaches a plateau set by the gas turbulent velocity at the probing scale $h_d$; this value represents an upper limit, as higher resolution would probe smaller scales and yield a lower dispersion.

    \item \textbf{$0.1<\mathrm{St}<1$ (marginal coupling):}
    This is the regime most favourable for planet formation through direct dust collapse. Dust trapping is strongest: spiral arms act as pressure maxima and potential minima, allowing $\rho_d/\rho_g$ to increase dramatically, up to values of order unity inside spirals.
    At the same time, dust excitation is minimised: dust motions within spiral arms become sufficiently correlated that the local velocity dispersion reaches its minimum. The combination of dust trapping and small $c_d$ makes this regime the most prone to two-fluid gravitational instability and collapse into bound clumps/planetary cores.

    \item \textbf{$1<\mathrm{St}<10$ (intermediate decoupling):}
    As $\mathrm{St}$ increases above unity, dust becomes progressively less coupled to the gas. Trapping in spiral arms weakens because drag is no longer effective at concentrating solids, while relative motions grow because spiral perturbations gravitationally ``kick'' dust onto more eccentric/inclined orbits. In this regime, drag still provides partial damping of these perturbations, but less efficiently as $\mathrm{St}$ increases.
    Consistently with \citet{booth16} and our results, the dust velocity dispersion increases approximately as $c_d/c_s \propto \sqrt{\mathrm{St}}$, reducing the likelihood of collapse.

    \item \textbf{$\mathrm{St}>10$ (weak coupling / ballistic limit):}
    For very large Stokes numbers, aerodynamic drag becomes negligible and dust dynamics is essentially ballistic \citep{gustavsson11,walmswell13}. Motions are controlled by the stellar potential plus gravitational perturbations from the spiral arms. The dust velocity dispersion saturates and is dominated by eccentric motions driven by spiral forcing.
    A simple estimate follows from $e \sim H/R$: the characteristic random velocity is $c_d \sim e\,v_k \sim (H/R)\,v_k \sim c_s$, i.e. $c_d/c_s = \mathcal{O}(1)$   \citep{walmswell13}. In this regime, trapping is inefficient and the large velocity dispersion strongly suppresses direct dust collapse.
\end{itemize}

\section{Conclusions}\label{sec_concl}
In this paper, we investigate the dynamics of dust particles in gravitationally unstable discs using 3D global simulations, exploring different degrees of aerodynamic coupling. We focus on the monodisperse case in Stokes number, meaning that we consider dust particles with the same Stokes number. We investigate three key quantities that determine the likelihood of planet formation at early stages of protoplanetary disc evolution, namely the dust-to-gas ratio, the collisional velocity between dust grains, and the velocity dispersion of the grains. We compare 2D \citep{booth16} and 3D SPH simulations, finding an overall very good agreement.

We find that the most relevant aerodynamic coupling regime occurs for St $\in(0.1, 1)$, where dust particles are strongly concentrated within spiral arms, their collisional velocities are expected to be minimal, and their velocity dispersion is very low. We show that, in this regime, dust particles are linearly unstable to the two-fluid gravitational instability, making dust collapse and planet formation possible. However, dust particles must first reach these sizes, which is not straightforward in such young discs. In this context, the collisional velocity provides insight into whether dust particles are likely to grow or fragment into smaller grains.

We find that the resolution requirements, both in terms of gas and dust particles, are very stringent in the low Stokes number regime, making predictions particularly uncertain in this case. Crucially, these limitations are not only due to the sampling of dust particles, but also to the ability of the simulations to resolve the turbulent cascade in the gas down to sufficiently small scales. Since dust relative velocities at low Stokes number are driven by gas turbulence, an accurate estimate of collisional velocities requires resolving gas fluctuations below the stopping length. If the turbulent cascade is truncated at larger scales, the inferred collisional velocities are biased and cannot be reliably interpreted. This work highlights the need for simulations that can simultaneously capture the small-scale structure of gas turbulence and the dynamics of tightly coupled dust, likely requiring local, high-resolution approaches. Our analysis of the gas power spectrum shows that SPH simulations can capture the turbulent cascade with high efficiency, reaching dissipation scales comparable to those achieved in high-resolution grid-based simulations. However, in a global disc setup the computational cost remains prohibitive, preventing a simultaneous resolution of both large and small scales.


\section*{Acknowledgements}
 {The authors thank the referee Hossam Aly for an insightful and thorough review of the paper, which significantly improved its quality. }
The authors thank Daniel Price for useful discussions. CL and CJC have been supported by the UK Science and Technology Research Council (STFC) via the consolidated grant ST/W000997/1.  CL acknowledges funding from the European Union’s Horizon 2020 research and innovation program under the Marie Skłodowska-Curie grant agreement N. 823823 (DUSTBUSTERS RISE project). RB acknowledges the Royal Society for their support via a University Research Fellowship.

The simulations presented in this work were performed using the DiRAC Data Intensive service at Leicester (DiAL3), operated by the University of Leicester IT Services, which is part of the STFC DiRAC HPC Facility (www.dirac.ac.uk) within the RAC large project DISCSIM IV and Cambridge Service for Data Driven Discovery (CSD3).

\section*{Data Availability}
Snapshots of the simulations, together with the setup files are available on this DOI \url{10.5281/zenodo.22041928}.



\bibliographystyle{mnras}
\bibliography{example} 

@ARTICLE{cossins09,
       author = {{Cossins}, Peter and {Lodato}, Giuseppe and {Clarke}, C.~J.},
        title = "{Characterizing the gravitational instability in cooling accretion discs}",
      journal = {\mnras},
         year = 2009,
        month = mar,
       volume = {393},
       number = {4},
        pages = {1157-1173},
          doi = {10.1111/j.1365-2966.2008.14275.x},
archivePrefix = {arXiv},
       eprint = {0811.3629},
 primaryClass = {astro-ph},
       adsurl = {https://ui.adsabs.harvard.edu/abs/2009MNRAS.393.1157C}
}

@ARTICLE{booth16,
       author = {{Booth}, Richard A. and {Clarke}, Cathie J.},
        title = "{Collision velocity of dust grains in self-gravitating protoplanetary discs}",
      journal = {\mnras},
         year = 2016,
        month = may,
       volume = {458},
       number = {3},
        pages = {2676-2693},
          doi = {10.1093/mnras/stw488},
archivePrefix = {arXiv},
       eprint = {1603.00029},
 primaryClass = {astro-ph.EP},
       adsurl = {https://ui.adsabs.harvard.edu/abs/2016MNRAS.458.2676B}
}

@ARTICLE{longarini23b,
       author = {{Longarini}, Cristiano and {Armitage}, Philip J. and {Lodato}, Giuseppe and {Price}, Daniel J. and {Ceppi}, Simone},
        title = "{The role of the drag force in the gravitational stability of dusty planet-forming disc - II. Numerical simulations}",
      journal = {\mnras},
         year = 2023,
        month = jul,
       volume = {522},
       number = {4},
        pages = {6217-6235},
          doi = {10.1093/mnras/stad1400},
archivePrefix = {arXiv},
       eprint = {2305.03659},
 primaryClass = {astro-ph.EP},
       adsurl = {https://ui.adsabs.harvard.edu/abs/2023MNRAS.522.6217L}
}

@ARTICLE{dipierro15,
       author = {{Dipierro}, Giovanni and {Pinilla}, Paola and {Lodato}, Giuseppe and {Testi}, Leonardo},
        title = "{Dust trapping by spiral arms in gravitationally unstable protostellar discs}",
      journal = {\mnras},
         year = 2015,
        month = jul,
       volume = {451},
       number = {1},
        pages = {974-986},
          doi = {10.1093/mnras/stv970},
archivePrefix = {arXiv},
       eprint = {1504.08099},
 primaryClass = {astro-ph.SR},
       adsurl = {https://ui.adsabs.harvard.edu/abs/2015MNRAS.451..974D}
}

@ARTICLE{rice04,
       author = {{Rice}, W.~K.~M. and {Lodato}, G. and {Pringle}, J.~E. and {Armitage}, P.~J. and {Bonnell}, I.~A.},
        title = "{Accelerated planetesimal growth in self-gravitating protoplanetary discs}",
      journal = {\mnras},
         year = 2004,
        month = dec,
       volume = {355},
       number = {2},
        pages = {543-552},
          doi = {10.1111/j.1365-2966.2004.08339.x},
archivePrefix = {arXiv},
       eprint = {astro-ph/0408390},
 primaryClass = {astro-ph},
       adsurl = {https://ui.adsabs.harvard.edu/abs/2004MNRAS.355..543R}
}

@ARTICLE{longarini23a,
       author = {{Longarini}, Cristiano and {Lodato}, Giuseppe and {Bertin}, Giuseppe and {Armitage}, Philip J.},
        title = "{The role of the drag force in the gravitational stability of dusty planet forming disc - I. Analytical theory}",
      journal = {\mnras},
         year = 2023,
        month = feb,
       volume = {519},
       number = {2},
        pages = {2017-2029},
          doi = {10.1093/mnras/stac3653},
archivePrefix = {arXiv},
       eprint = {2212.04986},
 primaryClass = {astro-ph.EP},
       adsurl = {https://ui.adsabs.harvard.edu/abs/2023MNRAS.519.2017L}
}

@ARTICLE{rowther24,
       author = {{Rowther}, Sahl and {Nealon}, Rebecca and {Meru}, Farzana and {Wurster}, James and {Aly}, Hossam and {Alexander}, Richard and {Rice}, Ken and {Booth}, Richard A.},
        title = "{The role of drag and gravity on dust concentration in a gravitationally unstable disc}",
      journal = {\mnras},
         year = 2024,
        month = feb,
       volume = {528},
       number = {2},
        pages = {2490-2500},
          doi = {10.1093/mnras/stae167},
archivePrefix = {arXiv},
       eprint = {2401.09380},
 primaryClass = {astro-ph.EP},
       adsurl = {https://ui.adsabs.harvard.edu/abs/2024MNRAS.528.2490R}
}

@ARTICLE{phantom,
       author = {{Price}, Daniel J. and {Wurster}, James and {Tricco}, Terrence S. and {Nixon}, Chris and {Toupin}, St{\'e}ven and {Pettitt}, Alex and {Chan}, Conrad and {Mentiplay}, Daniel and {Laibe}, Guillaume and {Glover}, Simon and {Dobbs}, Clare and {Nealon}, Rebecca and {Liptai}, David and {Worpel}, Hauke and {Bonnerot}, Cl{\'e}ment and {Dipierro}, Giovanni and {Ballabio}, Giulia and {Ragusa}, Enrico and {Federrath}, Christoph and {Iaconi}, Roberto and {Reichardt}, Thomas and {Forgan}, Duncan and {Hutchison}, Mark and {Constantino}, Thomas and {Ayliffe}, Ben and {Hirsh}, Kieran and {Lodato}, Giuseppe},
        title = "{Phantom: A Smoothed Particle Hydrodynamics and Magnetohydrodynamics Code for Astrophysics}",
      journal = {\pasa},
         year = 2018,
        month = sep,
       volume = {35},
          eid = {e031},
        pages = {e031},
          doi = {10.1017/pasa.2018.25},
archivePrefix = {arXiv},
       eprint = {1702.03930},
 primaryClass = {astro-ph.IM},
       adsurl = {https://ui.adsabs.harvard.edu/abs/2018PASA...35...31P}
}

@ARTICLE{monaghan2fl,
       author = {{Monaghan}, J.~J. and {Kocharyan}, A.},
        title = "{SPH simulation of multi-phase flow}",
      journal = {Computer Physics Communications},
         year = 1995,
        month = may,
       volume = {87},
       number = {1-2},
        pages = {225-235},
          doi = {10.1016/0010-4655(94)00174-Z},
       adsurl = {https://ui.adsabs.harvard.edu/abs/1995CoPhC..87..225M}
}

@ARTICLE{ceppi23,
       author = {{Ceppi}, Simone and {Longarini}, Cristiano and {Lodato}, Giuseppe and {Cuello}, Nicol{\'a}s and {Lubow}, Stephen H.},
        title = "{Precession and polar alignment of accretion discs in triple (or multiple) stellar systems}",
      journal = {\mnras},
         year = 2023,
        month = apr,
       volume = {520},
       number = {4},
        pages = {5817-5827},
          doi = {10.1093/mnras/stad444},
archivePrefix = {arXiv},
       eprint = {2302.03411},
 primaryClass = {astro-ph.EP},
       adsurl = {https://ui.adsabs.harvard.edu/abs/2023MNRAS.520.5817C}
}

@ARTICLE{prasad25,
       author = {{Prasad}, Vasundhara R. and {Longarini}, Cristiano and {Clarke}, Cathie J.},
        title = "{Dust trapping in protoplanetary discs after stellar flybys}",
      journal = {\mnras},
         year = 2025,
        month = oct,
       volume = {543},
       number = {2},
        pages = {1798-1815},
          doi = {10.1093/mnras/staf1579},
archivePrefix = {arXiv},
       eprint = {2509.11909},
 primaryClass = {astro-ph.EP},
       adsurl = {https://ui.adsabs.harvard.edu/abs/2025MNRAS.543.1798P}
}

@ARTICLE{nealon20,
       author = {{Nealon}, Rebecca and {Price}, Daniel J. and {Pinte}, Christophe},
        title = "{Rocking shadows in broken circumbinary discs}",
      journal = {\mnras},
         year = 2020,
        month = mar,
       volume = {493},
       number = {1},
        pages = {L143-L147},
          doi = {10.1093/mnrasl/slaa026},
archivePrefix = {arXiv},
       eprint = {2002.02983},
 primaryClass = {astro-ph.EP},
       adsurl = {https://ui.adsabs.harvard.edu/abs/2020MNRAS.493L.143N}
}

@ARTICLE{aly24,
       author = {{Aly}, Hossam and {Nealon}, Rebecca and {Gonzalez}, Jean-Fran{\c{c}}ois},
        title = "{WInDI: a Warp-Induced Dust Instability in protoplanetary discs}",
      journal = {\mnras},
         year = 2024,
        month = jan,
       volume = {527},
       number = {3},
        pages = {4777-4789},
          doi = {10.1093/mnras/stad3494},
archivePrefix = {arXiv},
       eprint = {2311.06182},
 primaryClass = {astro-ph.EP},
       adsurl = {https://ui.adsabs.harvard.edu/abs/2024MNRAS.527.4777A}
}

@ARTICLE{price20,
       author = {{Price}, Daniel J. and {Laibe}, Guillaume},
        title = "{A solution to the overdamping problem when simulating dust-gas mixtures with smoothed particle hydrodynamics}",
      journal = {\mnras},
         year = 2020,
        month = jul,
       volume = {495},
       number = {4},
        pages = {3929-3934},
          doi = {10.1093/mnras/staa1366},
archivePrefix = {arXiv},
       eprint = {2005.06562},
 primaryClass = {astro-ph.IM},
       adsurl = {https://ui.adsabs.harvard.edu/abs/2020MNRAS.495.3929P}
}

@ARTICLE{gammie01,
       author = {{Gammie}, Charles F.},
        title = "{Nonlinear Outcome of Gravitational Instability in Cooling, Gaseous Disks}",
      journal = {\apj},
         year = 2001,
        month = may,
       volume = {553},
       number = {1},
        pages = {174-183},
          doi = {10.1086/320631},
archivePrefix = {arXiv},
       eprint = {astro-ph/0101501},
 primaryClass = {astro-ph},
       adsurl = {https://ui.adsabs.harvard.edu/abs/2001ApJ...553..174G}
}

@ARTICLE{lodato04,
       author = {{Lodato}, G. and {Rice}, W.~K.~M.},
        title = "{Testing the locality of transport in self-gravitating accretion discs}",
      journal = {\mnras},
         year = 2004,
        month = jun,
       volume = {351},
       number = {2},
        pages = {630-642},
          doi = {10.1111/j.1365-2966.2004.07811.x},
archivePrefix = {arXiv},
       eprint = {astro-ph/0403185},
 primaryClass = {astro-ph},
       adsurl = {https://ui.adsabs.harvard.edu/abs/2004MNRAS.351..630L}
}

@ARTICLE{bethune21,
       author = {{B{\'e}thune}, William and {Latter}, Henrik and {Kley}, Wilhelm},
        title = "{Spiral structures in gravito-turbulent gaseous disks}",
      journal = {\aap},
         year = 2021,
        month = jun,
       volume = {650},
          eid = {A49},
        pages = {A49},
          doi = {10.1051/0004-6361/202040094},
archivePrefix = {arXiv},
       eprint = {2102.00775},
 primaryClass = {astro-ph.EP},
       adsurl = {https://ui.adsabs.harvard.edu/abs/2021A&A...650A..49B}
}

@ARTICLE{deng17,
       author = {{Deng}, Hongping and {Mayer}, Lucio and {Meru}, Farzana},
        title = "{Convergence of the Critical Cooling Rate for Protoplanetary Disk Fragmentation Achieved: The Key Role of Numerical Dissipation of Angular Momentum}",
      journal = {\apj},
         year = 2017,
        month = sep,
       volume = {847},
       number = {1},
          eid = {43},
        pages = {43},
          doi = {10.3847/1538-4357/aa872b},
archivePrefix = {arXiv},
       eprint = {1706.00417},
 primaryClass = {astro-ph.EP},
       adsurl = {https://ui.adsabs.harvard.edu/abs/2017ApJ...847...43D}
}

@ARTICLE{andrews18,
       author = {{Andrews}, Sean M. and {Huang}, Jane and {P{\'e}rez}, Laura M. and {Isella}, Andrea and {Dullemond}, Cornelis P. and {Kurtovic}, Nicol{\'a}s T. and {Guzm{\'a}n}, Viviana V. and {Carpenter}, John M. and {Wilner}, David J. and {Zhang}, Shangjia and {Zhu}, Zhaohuan and {Birnstiel}, Tilman and {Bai}, Xue-Ning and {Benisty}, Myriam and {Hughes}, A. Meredith and {{\"O}berg}, Karin I. and {Ricci}, Luca},
        title = "{The Disk Substructures at High Angular Resolution Project (DSHARP). I. Motivation, Sample, Calibration, and Overview}",
      journal = {\apjl},
         year = 2018,
        month = dec,
       volume = {869},
       number = {2},
          eid = {L41},
        pages = {L41},
          doi = {10.3847/2041-8213/aaf741},
archivePrefix = {arXiv},
       eprint = {1812.04040},
 primaryClass = {astro-ph.SR},
       adsurl = {https://ui.adsabs.harvard.edu/abs/2018ApJ...869L..41A}
}

@ARTICLE{segura-cox20,
       author = {{Segura-Cox}, Dominique M. and {Schmiedeke}, Anika and {Pineda}, Jaime E. and {Stephens}, Ian W. and {Fern{\'a}ndez-L{\'o}pez}, Manuel and {Looney}, Leslie W. and {Caselli}, Paola and {Li}, Zhi-Yun and {Mundy}, Lee G. and {Kwon}, Woojin and {Harris}, Robert J.},
        title = "{Four annular structures in a protostellar disk less than 500,000 years old}",
      journal = {\nat},
         year = 2020,
        month = oct,
       volume = {586},
       number = {7828},
        pages = {228-231},
          doi = {10.1038/s41586-020-2779-6},
archivePrefix = {arXiv},
       eprint = {2010.03657},
 primaryClass = {astro-ph.EP},
       adsurl = {https://ui.adsabs.harvard.edu/abs/2020Natur.586..228S}
}

@ARTICLE{sheehan18,
       author = {{Sheehan}, Patrick D. and {Eisner}, Josh A.},
        title = "{Multiple Gaps in the Disk of the Class I Protostar GY 91}",
      journal = {\apj},
         year = 2018,
        month = apr,
       volume = {857},
       number = {1},
          eid = {18},
        pages = {18},
          doi = {10.3847/1538-4357/aaae65},
archivePrefix = {arXiv},
       eprint = {1803.02847},
 primaryClass = {astro-ph.SR},
       adsurl = {https://ui.adsabs.harvard.edu/abs/2018ApJ...857...18S}
}

@ARTICLE{pollack96,
       author = {{Pollack}, James B. and {Hubickyj}, Olenka and {Bodenheimer}, Peter and {Lissauer}, Jack J. and {Podolak}, Morris and {Greenzweig}, Yuval},
        title = "{Formation of the Giant Planets by Concurrent Accretion of Solids and Gas}",
      journal = {\icarus},
         year = 1996,
        month = nov,
       volume = {124},
       number = {1},
        pages = {62-85},
          doi = {10.1006/icar.1996.0190},
       adsurl = {https://ui.adsabs.harvard.edu/abs/1996Icar..124...62P}
}

@ARTICLE{boss97,
       author = {{Boss}, A.~P.},
        title = "{Giant planet formation by gravitational instability.}",
      journal = {Science},
         year = 1997,
        month = jan,
       volume = {276},
        pages = {1836-1839},
          doi = {10.1126/science.276.5320.1836},
       adsurl = {https://ui.adsabs.harvard.edu/abs/1997Sci...276.1836B}
}

@ARTICLE{kratter16,
       author = {{Kratter}, Kaitlin and {Lodato}, Giuseppe},
        title = "{Gravitational Instabilities in Circumstellar Disks}",
      journal = {\araa},
         year = 2016,
        month = sep,
       volume = {54},
        pages = {271-311},
          doi = {10.1146/annurev-astro-081915-023307},
archivePrefix = {arXiv},
       eprint = {1603.01280},
 primaryClass = {astro-ph.SR},
       adsurl = {https://ui.adsabs.harvard.edu/abs/2016ARA&A..54..271K}
}

@ARTICLE{rice06,
       author = {{Rice}, W.~K.~M. and {Lodato}, G. and {Pringle}, J.~E. and {Armitage}, P.~J. and {Bonnell}, I.~A.},
        title = "{Planetesimal formation via fragmentation in self-gravitating protoplanetary discs}",
      journal = {\mnras},
         year = 2006,
        month = oct,
       volume = {372},
       number = {1},
        pages = {L9-L13},
          doi = {10.1111/j.1745-3933.2006.00215.x},
archivePrefix = {arXiv},
       eprint = {astro-ph/0607268},
 primaryClass = {astro-ph},
       adsurl = {https://ui.adsabs.harvard.edu/abs/2006MNRAS.372L...9R}
}

@ARTICLE{rice25,
       author = {{Rice}, Ken and {Baehr}, Hans and {Young}, Alison K. and {Booth}, Richard and {Rowther}, Sahl and {Meru}, Farzana and {Hall}, Cassandra and {Koval}, Adam},
        title = "{Dust density enhancements and the direct formation of planetary cores in gravitationally unstable discs}",
      journal = {\mnras},
         year = 2025,
        month = jun,
       volume = {539},
       number = {4},
        pages = {3421-3435},
          doi = {10.1093/mnras/staf714},
archivePrefix = {arXiv},
       eprint = {2505.00363},
 primaryClass = {astro-ph.EP},
       adsurl = {https://ui.adsabs.harvard.edu/abs/2025MNRAS.539.3421R}
}

@ARTICLE{birnstiel16,
       author = {{Birnstiel}, T. and {Fang}, M. and {Johansen}, A.},
        title = "{Dust Evolution and the Formation of Planetesimals}",
      journal = {\ssr},
         year = 2016,
        month = dec,
       volume = {205},
       number = {1-4},
        pages = {41-75},
          doi = {10.1007/s11214-016-0256-1},
archivePrefix = {arXiv},
       eprint = {1604.02952},
 primaryClass = {astro-ph.SR},
       adsurl = {https://ui.adsabs.harvard.edu/abs/2016SSRv..205...41B}
}

@ARTICLE{Leedham26prep,
       author = {{Leedham C. in prep}, C. et al.},
        title = "{In prep}",
      journal = {\mnras},
         year = 2026
}

@ARTICLE{walmswell13,
       author = {{Walmswell}, Joe and {Clarke}, Cathie and {Cossins}, Peter},
        title = "{The evolution of planetesimal swarms in self-gravitating protoplanetary discs}",
      journal = {\mnras},
         year = 2013,
        month = may,
       volume = {431},
       number = {2},
        pages = {1903-1913},
          doi = {10.1093/mnras/stt314},
archivePrefix = {arXiv},
       eprint = {1302.7216},
 primaryClass = {astro-ph.EP},
       adsurl = {https://ui.adsabs.harvard.edu/abs/2013MNRAS.431.1903W}
}

@ARTICLE{gustavsson11,
       author = {{Gustavsson}, K. and {Mehlig}, B.},
        title = "{Distribution of relative velocities in turbulent aerosols}",
      journal = {\pre},
         year = 2011,
        month = oct,
       volume = {84},
       number = {4},
          eid = {045304},
        pages = {045304},
          doi = {10.1103/PhysRevE.84.045304},
archivePrefix = {arXiv},
       eprint = {1012.1789},
 primaryClass = {physics.flu-dyn},
       adsurl = {https://ui.adsabs.harvard.edu/abs/2011PhRvE..84d5304G}
}

@ARTICLE{baehr22,
       author = {{Baehr}, Hans and {Zhu}, Zhaohuan and {Yang}, Chao-Chin},
        title = "{Direct Formation of Planetary Embryos in Self-gravitating Disks}",
      journal = {\apj},
         year = 2022,
        month = jul,
       volume = {933},
       number = {1},
          eid = {100},
        pages = {100},
          doi = {10.3847/1538-4357/ac7228},
archivePrefix = {arXiv},
       eprint = {2204.13310},
 primaryClass = {astro-ph.EP},
       adsurl = {https://ui.adsabs.harvard.edu/abs/2022ApJ...933..100B}
}

@ARTICLE{baehr21,
       author = {{Baehr}, Hans and {Zhu}, Zhaohuan},
        title = "{Particle Dynamics in 3D Self-gravitating Disks. I. Spirals}",
      journal = {\apj},
         year = 2021,
        month = mar,
       volume = {909},
       number = {2},
          eid = {135},
        pages = {135},
          doi = {10.3847/1538-4357/abddb3},
archivePrefix = {arXiv},
       eprint = {2101.01888},
 primaryClass = {astro-ph.EP},
       adsurl = {https://ui.adsabs.harvard.edu/abs/2021ApJ...909..135B}
}

@ARTICLE{gibbons12,
       author = {{Gibbons}, P.~G. and {Rice}, W.~K.~M. and {Mamatsashvili}, G.~R.},
        title = "{Planetesimal formation in self-gravitating discs}",
      journal = {\mnras},
         year = 2012,
        month = oct,
       volume = {426},
       number = {2},
        pages = {1444-1454},
          doi = {10.1111/j.1365-2966.2012.21731.x},
archivePrefix = {arXiv},
       eprint = {1207.4677},
 primaryClass = {astro-ph.EP},
       adsurl = {https://ui.adsabs.harvard.edu/abs/2012MNRAS.426.1444G}
}

@ARTICLE{gibbons14,
       author = {{Gibbons}, P.~G. and {Mamatsashvili}, G.~R. and {Rice}, W.~K.~M.},
        title = "{Planetesimal formation in self-gravitating discs - the effects of particle self-gravity and back-reaction}",
      journal = {\mnras},
         year = 2014,
        month = jul,
       volume = {442},
       number = {1},
        pages = {361-371},
          doi = {10.1093/mnras/stu809},
archivePrefix = {arXiv},
       eprint = {1404.6953},
 primaryClass = {astro-ph.EP},
       adsurl = {https://ui.adsabs.harvard.edu/abs/2014MNRAS.442..361G}
}

@ARTICLE{gibbons15,
       author = {{Gibbons}, P.~G. and {Mamatsashvili}, G.~R. and {Rice}, W.~K.~M.},
        title = "{Planetesimal formation in self-gravitating discs - dust trapping by vortices}",
      journal = {\mnras},
         year = 2015,
        month = nov,
       volume = {453},
       number = {4},
        pages = {4232-4243},
          doi = {10.1093/mnras/stv1766},
archivePrefix = {arXiv},
       eprint = {1508.02879},
 primaryClass = {astro-ph.EP},
       adsurl = {https://ui.adsabs.harvard.edu/abs/2015MNRAS.453.4232G}
}

@ARTICLE{booth19,
       author = {{Booth}, Richard A. and {Clarke}, Cathie J.},
        title = "{Characterizing gravito-turbulence in 3D: turbulent properties and stability against fragmentation}",
      journal = {\mnras},
         year = 2019,
        month = mar,
       volume = {483},
       number = {3},
        pages = {3718-3729},
          doi = {10.1093/mnras/sty3340},
archivePrefix = {arXiv},
       eprint = {1812.05644},
 primaryClass = {astro-ph.EP},
       adsurl = {https://ui.adsabs.harvard.edu/abs/2019MNRAS.483.3718B}
}

@ARTICLE{volk80,
       author = {{Voelk}, H.~J. and {Jones}, F.~C. and {Morfill}, G.~E. and {Roeser}, S.},
        title = "{Collisions between Grains in a Turbulent Gas}",
      journal = {\aap},
         year = 1980,
        month = may,
       volume = {85},
       number = {3},
        pages = {316-325},
       adsurl = {https://ui.adsabs.harvard.edu/abs/1980A&A....85..316V}
}

@ARTICLE{ormel07,
       author = {{Ormel}, C.~W. and {Cuzzi}, J.~N.},
        title = "{Closed-form expressions for particle relative velocities induced by turbulence}",
      journal = {\aap},
         year = 2007,
        month = may,
       volume = {466},
       number = {2},
        pages = {413-420},
          doi = {10.1051/0004-6361:20066899},
archivePrefix = {arXiv},
       eprint = {astro-ph/0702303},
 primaryClass = {astro-ph},
       adsurl = {https://ui.adsabs.harvard.edu/abs/2007A&A...466..413O}
}

@ARTICLE{dubrulle95,
       author = {{Dubrulle}, B. and {Morfill}, G. and {Sterzik}, M.},
        title = "{The dust subdisk in the protoplanetary nebula.}",
      journal = {\icarus},
         year = 1995,
        month = apr,
       volume = {114},
       number = {2},
        pages = {237-246},
          doi = {10.1006/icar.1995.1058},
       adsurl = {https://ui.adsabs.harvard.edu/abs/1995Icar..114..237D}
}

@ARTICLE{riols18,
       author = {{Riols}, A. and {Latter}, H.},
        title = "{Spiral density waves and vertical circulation in protoplanetary discs}",
      journal = {\mnras},
         year = 2018,
        month = jun,
       volume = {476},
       number = {4},
        pages = {5115-5126},
          doi = {10.1093/mnras/sty460},
archivePrefix = {arXiv},
       eprint = {1802.06620},
 primaryClass = {astro-ph.EP},
       adsurl = {https://ui.adsabs.harvard.edu/abs/2018MNRAS.476.5115R}
}

@ARTICLE{gundlach15,
       author = {{Gundlach}, B. and {Blum}, J.},
        title = "{The Stickiness of Micrometer-sized Water-ice Particles}",
      journal = {\apj},
         year = 2015,
        month = jan,
       volume = {798},
       number = {1},
          eid = {34},
        pages = {34},
          doi = {10.1088/0004-637X/798/1/34},
archivePrefix = {arXiv},
       eprint = {1410.7199},
 primaryClass = {astro-ph.EP},
       adsurl = {https://ui.adsabs.harvard.edu/abs/2015ApJ...798...34G}
}

@ARTICLE{birnstiel24,
       author = {{Birnstiel}, Tilman},
        title = "{Dust Growth and Evolution in Protoplanetary Disks}",
      journal = {\araa},
         year = 2024,
        month = sep,
       volume = {62},
       number = {1},
        pages = {157-202},
          doi = {10.1146/annurev-astro-071221-052705},
archivePrefix = {arXiv},
       eprint = {2312.13287},
 primaryClass = {astro-ph.EP},
       adsurl = {https://ui.adsabs.harvard.edu/abs/2024ARA&A..62..157B}
}

@ARTICLE{guttler10,
       author = {{G{\"u}ttler}, C. and {Blum}, J. and {Zsom}, A. and {Ormel}, C.~W. and {Dullemond}, C.~P.},
        title = "{The outcome of protoplanetary dust growth: pebbles, boulders, or planetesimals?. I. Mapping the zoo of laboratory collision experiments}",
      journal = {\aap},
         year = 2010,
        month = apr,
       volume = {513},
          eid = {A56},
        pages = {A56},
          doi = {10.1051/0004-6361/200912852},
archivePrefix = {arXiv},
       eprint = {0910.4251},
 primaryClass = {astro-ph.EP},
       adsurl = {https://ui.adsabs.harvard.edu/abs/2010A&A...513A..56G}
}

@ARTICLE{gartner17,
       author = {{G{\"a}rtner}, S. and {Gundlach}, B. and {Headen}, T.~F. and {Ratte}, J. and {Oesert}, J. and {Gorb}, S.~N. and {Youngs}, T.~G.~A. and {Bowron}, D.~T. and {Blum}, J. and {Fraser}, H.~J.},
        title = "{Micrometer-sized Water Ice Particles for Planetary Science Experiments: Influence of Surface Structure on Collisional Properties}",
      journal = {\apj},
         year = 2017,
        month = oct,
       volume = {848},
       number = {2},
          eid = {96},
        pages = {96},
          doi = {10.3847/1538-4357/aa8c7f},
archivePrefix = {arXiv},
       eprint = {1710.02074},
 primaryClass = {astro-ph.EP},
       adsurl = {https://ui.adsabs.harvard.edu/abs/2017ApJ...848...96G}
}

@ARTICLE{musiolik19,
       author = {{Musiolik}, Grzegorz and {Wurm}, Gerhard},
        title = "{Contacts of Water Ice in Protoplanetary Disks{\textemdash}Laboratory Experiments}",
      journal = {\apj},
         year = 2019,
        month = mar,
       volume = {873},
       number = {1},
          eid = {58},
        pages = {58},
          doi = {10.3847/1538-4357/ab0428},
archivePrefix = {arXiv},
       eprint = {1902.08503},
 primaryClass = {astro-ph.EP},
       adsurl = {https://ui.adsabs.harvard.edu/abs/2019ApJ...873...58M}
}

@ARTICLE{cullen10,
       author = {{Cullen}, Lee and {Dehnen}, Walter},
        title = "{Inviscid smoothed particle hydrodynamics}",
      journal = {\mnras},
         year = 2010,
        month = oct,
       volume = {408},
       number = {2},
        pages = {669-683},
          doi = {10.1111/j.1365-2966.2010.17158.x},
archivePrefix = {arXiv},
       eprint = {1006.1524},
 primaryClass = {astro-ph.IM},
       adsurl = {https://ui.adsabs.harvard.edu/abs/2010MNRAS.408..669C}
}

@ARTICLE{price12,
       author = {{Price}, Daniel J.},
        title = "{Resolving high Reynolds numbers in smoothed particle hydrodynamics simulations of subsonic turbulence}",
      journal = {\mnras},
         year = 2012,
        month = feb,
       volume = {420},
       number = {1},
        pages = {L33-L37},
          doi = {10.1111/j.1745-3933.2011.01187.x},
archivePrefix = {arXiv},
       eprint = {1111.1255},
 primaryClass = {astro-ph.CO},
       adsurl = {https://ui.adsabs.harvard.edu/abs/2012MNRAS.420L..33P}
}





\bsp	
\label{lastpage}
\end{document}